\documentclass{aa}
\usepackage[utf8]{inputenc}
\usepackage[T1]{fontenc}
\usepackage[english]{babel}
\usepackage{graphicx}
\usepackage{amsmath}
\usepackage{txfonts}
\usepackage{booktabs}
\usepackage{placeins}
\usepackage[hypertexnames=false,hyperfootnotes=false]{hyperref}
\hypersetup{hidelinks,pdftitle={Model-independent calibration of gamma-ray bursts with Bayesian neural networks},pdfauthor={P. Caterino et al.}}

\begin{document}
\title{Model-independent calibration of gamma-ray bursts with Bayesian neural networks}
\titlerunning{Model-independent calibration of GRBs}
\author{P. Caterino\inst{1}
\and M. Demianski\inst{2}
\and E. Piedipalumbo\inst{1,3}\corrauth{ester@na.infn.it}
\and L. Amati\inst{4}
\and D. De Cicco\inst{1}
\and M. Brescia\inst{1,5}}
\authorrunning{P. Caterino et al.}
\institute{Dipartimento di Fisica, Universit\`a degli Studi di Napoli Federico II, Compl. Univ. Monte S. Angelo, 80126 Naples, Italy
\and Institute for Theoretical Physics, University of Warsaw, Pasteura 5, 02-093 Warsaw, Poland
\and INFN, Sez. di Napoli, Compl. Univ. Monte S. Angelo, Edificio 6, via Cinthia, 80126 Naples, Italy
\and INAF-IASF, Sezione di Bologna, via Gobetti 101, 40129 Bologna, Italy
\and INAF--Osservatorio Astronomico di Capodimonte, Via Moiariello 16, 80131 Naples, Italy}
\date{}
\abstract
{Gamma-ray bursts (GRBs) extend cosmological distance measurements beyond the redshift range densely sampled by Type Ia supernovae, but their standardization requires reliable calibration and uncertainty estimates.}
{We calibrate the Amati correlation without assuming a background cosmological model, assess its stability, and examine the cosmological leverage of current and future GRB Hubble diagrams.}
{We train a Bayesian neural network on Pantheon+ supernovae, assess and recalibrate its predictive uncertainties, and calibrate the correlation for GRBs at $z\leq1.5$ using the D'Agostini and Reichart likelihoods. The resulting Hubble diagram is combined with supernovae, observational Hubble measurements, and baryon acoustic oscillations from the Dark Energy Spectroscopic Instrument. Local-regression, covariance, leave-three-out, redshift-evolution, and Efron--Petrosian diagnostics test the standardization assumptions. We also analyse idealized future catalogues of 1500 standardized GRBs.}
{The calibrated diagram extends to $z\simeq9.4$. Covariance and leave-three-out diagnostics produce small changes in the cosmological constraints. A parametric evolution coefficient is consistent with zero, whereas the truncation-corrected diagnostic indicates marginal isotropic-energy evolution that can occur along the Amati relation. Dark-energy extensions improve the phenomenological fit relative to flat $\Lambda$CDM: the Akaike criterion favours the Chevallier--Polarski--Linder parametrization, while the Bayesian criterion favours constant-$w$ dark energy. Future-catalogue forecasts show strong conditional model-discrimination power under the adopted uncertainty model.}
{GRBs provide a complementary high-redshift extension of the supernova distance ladder. Present results do not establish an evolving dark-energy equation of state. Larger homogeneous samples are promising, but realistic selection functions and calibration systematics remain essential.}
\maketitle
\nolinenumbers
\section{Introduction}\label{sec:introduction}
The accelerated expansion of the Universe is usually described in terms of dark energy. Precision measurements from Type Ia supernovae (SNIa), baryon acoustic oscillations (BAO), and the cosmic microwave background have established a successful reference cosmology, but also revealed tensions that could reflect unrecognized systematics or departures from flat $\Lambda$CDM. Recent analyses have renewed interest in a redshift-dependent dark-energy equation of state \citep{divalentino25,adame25}. Complementary probes spanning a wider redshift interval are therefore valuable both for testing these indications and for reducing parameter degeneracies.

Gamma-ray bursts (GRBs) are detectable to much higher redshifts than the densely sampled supernova Hubble diagram. Their isotropic-equivalent energies can approach $10^{54}$ erg, and the present sample reaches $z\simeq9.4$. Their prompt hard-X-ray and gamma-ray emission is little affected by dust. GRBs are not standard candles, however: their radiated energies span several orders of magnitude. Their cosmological use depends on empirical relations between distance-dependent quantities and rest-frame observables. Here we study the Amati correlation \citep{amati02,amati06},
\begin{equation}\label{correlation}
\log_{10}\left(\frac{E_{\rm iso}}{\rm erg}\right)
=b+a\log_{10}\left(\frac{E_{\rm p,i}}{300\,{\rm keV}}\right),
\end{equation}
where $E_{\rm p,i}=(1+z)E_{\rm p,obs}$ is the rest-frame spectral peak energy and
\begin{equation}\label{eqEiso}
E_{\rm iso}=\frac{4\pi d_L^2(z)S_{\rm bolo}}{1+z}
\end{equation}
is the isotropic-equivalent energy inferred from the bolometric fluence $S_{\rm bolo}$ and luminosity distance $d_L$. The slope $a$, intercept $b$, and intrinsic scatter must be calibrated before the relation can yield distances.

The scarcity of very nearby GRBs makes this calibration less direct than for SNIa. Assuming a cosmological model to determine $E_{\rm iso}$ and subsequently using the fitted relation to constrain that model introduces circularity. Both simultaneous self-calibration and low-redshift anchoring techniques have been proposed \citep{amati08,mci21,amati19,muccino21,izzo15,wdl15,wei10}. Low-redshift methods avoid choosing a background expansion model but inherit the distance scale and statistical information of their calibrating sample; their extrapolation also assumes that the relation remains applicable at higher redshift.

We reconstruct the Pantheon+ distance-modulus relation with a Bayesian neural network (BNN), explicitly assess its predictive uncertainties, and recalibrate their empirical coverage. The reconstruction anchors the Amati relation for GRBs at $z\leq1.5$. We compare the D'Agostini and Reichart likelihoods, construct a high-redshift Hubble diagram, and combine it with supernova, Hubble-parameter, and BAO measurements. Controlled local-regression tests quantify covariance inheritance, sensitivity to the highest-redshift supernovae, and possible redshift evolution. A complementary Efron--Petrosian analysis addresses an effective fluence truncation.

Here ‘model-independent’ means that no parametric form of the background expansion history is assumed during calibration; the resulting distances nevertheless inherit the Pantheon+ distance scale and its associated statistical and systematic information.

GRBs are used here as complementary high-redshift tracers rather than as competitors to the present precision of low-redshift probes. Their long redshift lever arm is also relevant to testing alternative expansion histories \citep{Demianski2020,Lusso25}. We finally assess the conditional cosmological reach of idealized future standardized-GRB samples inspired by THESEUS. Sections~\ref{sec:data}--\ref{sec:calibration} present the data, reconstruction, and GRB distances; Sects.~\ref{sec:cosmology}--\ref{sec:robustness} discuss cosmological constraints and robustness; Sect.~\ref{sec:theseus} gives the forecasts. Technical diagnostics and extended numerical results are collected in the appendices.

\section{Data and distance reconstruction}\label{sec:data}
\subsection{The GRB and supernova samples}
\begin{figure*}
\centering
\includegraphics[width=0.48\textwidth]{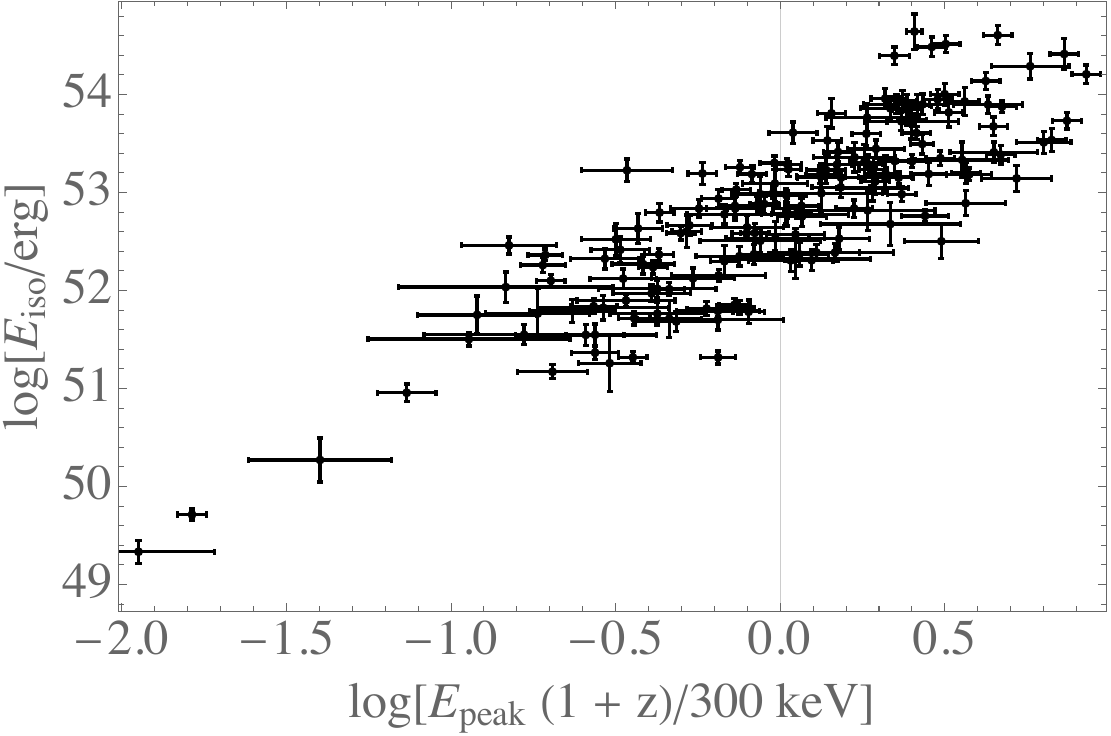}\hfill
\includegraphics[width=0.48\textwidth]{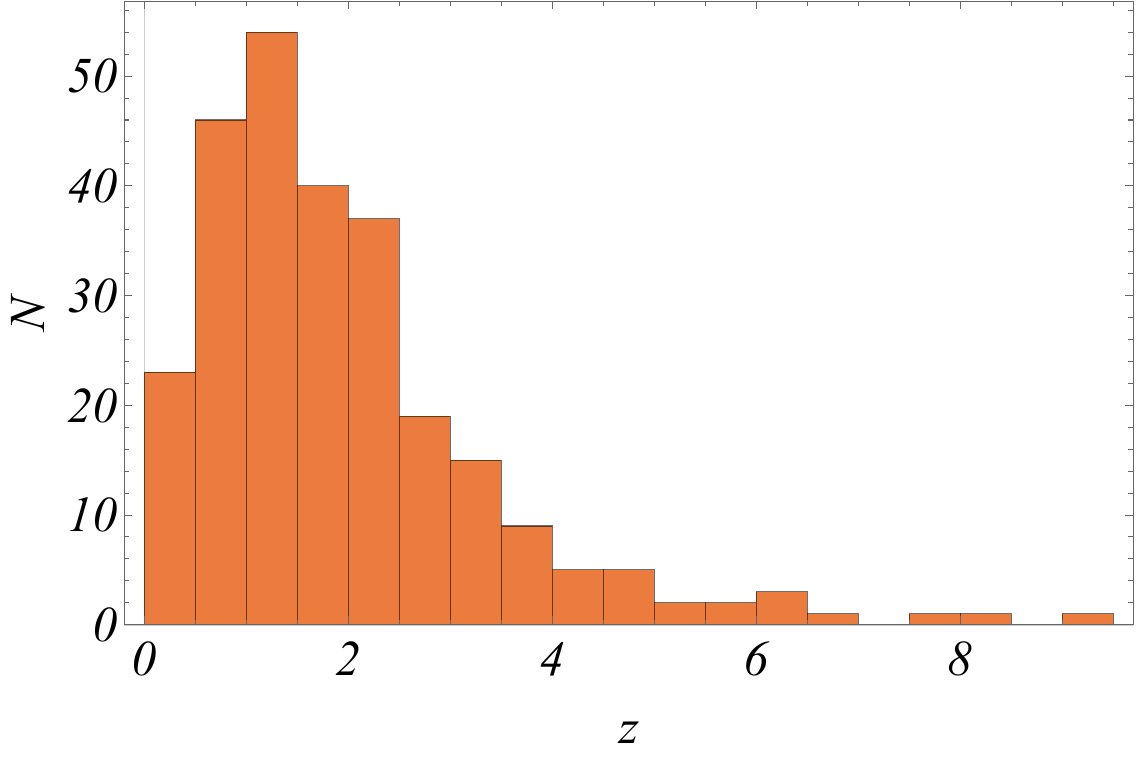}
\caption{The 264-GRB sample. Left: spectral peak energy and isotropic-equivalent energy. Right: redshift distribution.}
\label{fig:grb-sample}
\end{figure*}

Our compilation contains 264 long GRBs with measured redshifts and prompt-emission spectra, observed through the end of 2022 by CGRO/BATSE, BeppoSAX/GRBM, HETE--2, Swift/BAT, Fermi/GBM, and Konus--WIND. The same sample was used by \citet{cogato24}. It updates earlier compilations using systematic spectral catalogues \citep{afg09,katsukura20,tsvetkova21} and GCN Circulars\footnote{\url{https://gcn.gsfc.nasa.gov/gcn3_archive.html}}. Figure~\ref{fig:grb-sample} shows its energy distribution and redshift coverage.

We require spectral exposure over at least two-thirds of the burst duration. Konus--WIND and Fermi/GBM measurements are prioritized because of their broad energy coverage; Suzaku measurements are excluded. Swift/BAT is used only when the preferred instruments are unavailable and the observed peak lies within the BAT band. Reported uncertainties below 10\% are raised to 10\%. The Band spectral model is preferred to a cut-off power law when available. These choices reduce known instrumental limitations but do not define a complete selection function. The bolometric correction integrates the fitted spectrum over the observer-frame interval corresponding to 1--$10^4$ keV in the rest frame.

The Amati relation was initially established for long GRBs and subsequently tested with different instruments \citep{amati02,ldg05,amati06}. Short GRBs generally do not follow the same relation; the subenergetic GRBs 980425 and 031203 are also known exceptions \citep{amati07,antonelli09}. Its extra-Poissonian dispersion must be included in the calibration and propagated to the distance errors \citep{amati08,afg09}. Cross-instrument comparisons and studies of BATSE bursts support broad consistency with the relation \citep{ghirlanda08,nava11}, but do not remove the need to test selection and evolution explicitly. Earlier redshift-binned and evolving-relation analyses provide useful context \citep{ad13,demianski21}; we reassess these issues in Sect.~\ref{sec:robustness}.

The distance anchor is Pantheon+ \citep{brout22}, with 1701 light curves of 1550 spectroscopically confirmed SNIa over $0.001<z<2.26$. Repeated observations enter as separate light curves, with the released statistical-plus-systematic covariance describing their dependence. The GRB calibration is restricted to $z\leq1.5$, where the supernova reconstruction remains empirically better constrained than at the sparsely populated high-redshift end. Thus the calibrated GRBs extend the Pantheon+ distance ladder; they do not independently determine its absolute normalization.

\subsection{Bayesian reconstruction and uncertainty calibration}\label{sec:bnn}
A BNN  is a probabilistic extension of neural networks that integrates Bayesian inference into the learning process, providing a useful approach to modeling uncertainties \citep{jospin22}. Unlike conventional neural networks, where weights are treated as fixed parameters optimized through deterministic training, a BNN assigns a prior probability distribution to each weight and then updates this distribution through a process of Bayesian inference.
Formally, given a dataset  $\mathcal{D} = \{x_i,y_i \}_{i=1}^n $ , a BNN defines a probabilistic model for the conditional distribution $ p(y|x,\mathcal{W})$ where $\mathcal{W}$ denotes the set of network weights.
A prior distribution $p(\mathcal{W})$ is placed over the weights, encapsulating the initial assumptions about the parameters. Bayesian inference is used to update this prior in light of the observed data, leading to a posterior distribution
\begin{equation}p(\mathcal{W}|\mathcal{D})  = \frac{p(\mathcal{D}|\mathcal{W})p(\mathcal{W})}{p(\mathcal{D})}\,,  \end{equation}
that reflects both prior knowledge and empirical evidence. The predictive distribution for a new data point $x^*$ is then obtained by marginalizing over the posterior, as expressed by
\begin{equation} p(y^*|x^*,\mathcal{D}) = \int p(y^*| x^*,\mathcal{W})p(\mathcal{W}|\mathcal{D}) d\,\mathcal{W}\,.  \end{equation}
However, calculating the exact posterior distribution is often impractical for deep neural networks due to their high-dimensional parameter space. To counter this, methods like variational inference (VI) and Markov chain Monte Carlo (MCMC) are used. MCMC methods can provide more accurate posterior samples but are computationally expensive, whereas VI offers a more efficient approximation at the cost of additional approximation error.

Among VI methods, Monte Carlo Dropout is efficient and scales well. Originally used as a regularization technique, MC Dropout can be interpreted as a form of approximate variational inference when applied at test time. In Monte Carlo Dropout, neurons are randomly deactivated with a probability, $1-p$, called the dropout rate. By randomly dropping units during both training and inference, the network effectively samples from a distribution over its weights.

Monte Carlo dropout approximates the posterior by performing multiple forward passes with different dropout masks applied to neurons. The masks independently retain each neuron with probability $p$ and drop it with probability $1-p$. This randomization effectively simulates different weight configurations for the network, allowing the model to capture the uncertainty by treating the resulting predictions as samples from the approximate posterior distribution
\begin{equation} p(y^*| x^*,\mathcal{D}) \approx \frac{1}{T} \sum_{t=1}^T p(y^*|x^*,\mathcal{W}_t)\,. \end{equation}
Here $T$ indicates the total number of  collected samples, and $\mathcal{W}_t$ the network weight sampled by using a different dropout mask at each forward pass.
Despite its computational efficiency, MC dropout has notable limitations. The dropout rate $p$ is a hyper-parameter that requires manual tuning, and a suboptimal choice can significantly degrade both model performance and the reliability of uncertainty estimates.
To address this challenge, we use Concrete Dropout (CD; \citealt{ghk17}), which treats the retention probability $p$ — the probability of keeping a neuron active — as a learnable parameter within a variational Bayesian framework.
Concrete Dropout samples binary masks from a continuous concrete distribution, defined for neuron  $i$  as:
\begin{equation}
z_i= \text{sigmoid}\bigg( \frac{1}{t}(\log{p}-\log{(1-p)} + \log{u}- \log{(1-u)})\bigg)\,\end{equation}
where $u \sim \text{Uniform}(0,1)$ is a uniform random variable and $t$ is the temperature parameter controlling the proximity of sampled values to binary $0$ and $1$.
CD enhances model performance and uncertainty calibration by optimizing layer-specific retention probabilities during training, unlike MC dropout's fixed rates.

While CD eliminates manual tuning of dropout rates, it introduces two hyper-parameters: the temperature $t$, which controls the smoothness of the dropout distribution, and a regularization length scale, which regularizes the learned retention probabilities to prevent overfitting. These are simpler to calibrate than MC dropout’s rate, which requires exhaustive grid searches across multiple values per layer.
$t$  is relatively insensitive and can be set to 0.1 for most tasks. For the prior length scale, \citet{ghk17} proposed a default value proportional to (1/N), where (N) is the size of the training dataset. In addition, the Concrete Dropout regularization depends on a model precision parameter $\tau$, which controls the strength of the variational regularization through its relation to the expected observation noise. In our case, $\tau$ was estimated from the inverse variance of the residuals of a best-fit $\Lambda$CDM model to the Pantheon+ dataset, yielding $\tau = 34$. However, exploratory tests performed over a  range of $\tau$ values ($[5,34]$) produced nearly identical predictive reconstructions and calibration metrics, indicating that the final results are largely insensitive to the precise choice of this hyperparameter in the investigated range.
The BNN used in this study is implemented using TensorFlow and TensorFlow Probability. The network receives the supernova redshift as input and consists of a normalization layer followed by two fully connected hidden layers with 300 swish-activated neurons. Concrete Dropout layers are applied after each hidden layer, enabling a variational approximation to the posterior distribution over the network weights while simultaneously learning the dropout probabilities.
The final dense layer contains two neurons that parameterize a Gaussian predictive distribution. The first output corresponds to the predicted distance modulus, $\mu(z)$, while the second output is transformed through a softplus activation to obtain a strictly positive predictive standard deviation, $\sigma_{\mathrm{pred}}(z)$. 

These quantities define the predictive distribution
\begin{equation}
p(y|z,\theta)=\mathcal{N}\left(\mu_\theta(z),\sigma_{\mathrm{pred}}^2(z)\right),
\end{equation}
which is implemented through TensorFlow Probability's \texttt{DistributionLambda} layer. The resulting model produces a predictive distribution characterized by a mean $\mu(z)$ and a learned heteroscedastic predictive variance $\sigma_{\mathrm{pred}}^2(z)$. Epistemic uncertainty is estimated independently through repeated Monte Carlo sampling of the Concrete Dropout posterior, and is subsequently combined with the predictive variance to obtain the total predictive uncertainty of the reconstruction.

The Pantheon+ dataset is then divided into training, calibration,
and test sets in a 70/15/15 ratio. The corresponding within-subset
covariance matrices are extracted from the full Pantheon+ covariance
matrix.
To ensure robust learning in the sparsely populated high-redshift regime, the three most distant supernovae $z>1.7$ are included in the training sample.
To assess the predictive calibration in the redshift regime relevant for GRB applications, supernovae with $z<0.02$ are excluded from the calibration and test samples. These nearby objects primarily anchor the local Hubble flow and constitute a highly over-represented population within the Pantheon+ compilation. Since the subsequent GRB calibration is performed at substantially higher redshifts, retaining these objects in the evaluation sets would disproportionately influence the calibration metrics while providing limited information about the predictive performance of the model in the regime of interest. The low-redshift supernovae are nevertheless retained in the training sample to constrain the distance-modulus relation at the lowest redshifts.
The strong correlations present in the Pantheon+ covariance model lead some of the covariance submatrices obtained after the train--calibration--test split to be nearly singular, resulting in numerical instabilities during matrix inversion and determinant evaluation. To ensure stable likelihood optimization, a small Tikhonov regularization term is added to each covariance matrix,
\begin{equation}
\mathbf{C}_{\mathrm{SN}}
\rightarrow
\mathbf{C}_{\mathrm{SN}}
+
\lambda \mathbf{I},
\end{equation}
with $\lambda = 10^{-6}$. This modification improves the conditioning of the covariance matrices while remaining several orders of magnitude smaller than the dominant covariance elements, thereby preserving the statistical content of the Pantheon+ uncertainty model.
The details of the likelihood construction and the training procedure are presented in the following section.
\paragraph{Full-covariance training likelihood.}
The Pantheon+ supernova compilation is characterized by a non-diagonal covariance matrix that accounts for both statistical and systematic correlations between observations. To properly incorporate this information into the BNN training procedure, we model the data using a multivariate Gaussian likelihood whose covariance combines the experimental covariance of the catalog with a learned heteroscedastic predictive variance.
For a set of $N$ supernovae, the BNN predicts both a mean distance modulus $\boldsymbol{\mu}_{\theta}(z)$ and a redshift-dependent predictive standard deviation $\boldsymbol{\sigma}_{\mathrm{pred}}$. The total covariance is defined as
\begin{equation}
\label{eq:bnn-nll}
\boldsymbol{\Sigma}_{\mathrm{tot}} = \mathbf{C}_{\mathrm{SN}} + \mathrm{diag}(\sigma_{\mathrm{pred}}^2),
\end{equation}
where $\mathbf{C}_{\mathrm{SN}}$ is the Pantheon+ covariance matrix and $\mathrm{diag}(\sigma_{\mathrm{pred}}^2)$ is a diagonal matrix containing the variance predicted by the network.

The model parameters are optimized by minimizing the negative log-likelihood (NLL)
\begin{equation}
\mathcal{L}_{\mathrm{NLL}} = \frac{1}{2} [(\mathbf{y}-\boldsymbol{\mu}_{\theta})^{T}
\boldsymbol{\Sigma}_{\mathrm{tot}}^{-1}
(\mathbf{y}-\boldsymbol{\mu}_{\theta}) +\ln |\boldsymbol{\Sigma}_{\mathrm{tot}}|]
\end{equation}
where $\mathbf{y}$ denotes the observed distance moduli. The first term corresponds to a Mahalanobis distance weighted by the full covariance structure of the dataset, while the logarithmic determinant acts as a normalization term and naturally regularizes the predicted variance. This prevents the network from trivially minimizing the residual term through an artificial inflation of the uncertainty estimates.

To accelerate convergence in the Pantheon+ redshift distribution, we employ an initial warmup phase lasting 5000 epochs. During this stage, the optimization objective is augmented with a weighted mean squared error term
\begin{equation}
\mathcal{L}_{\mathrm{warm}}
=
\mathcal{L}_{\mathrm{NLL}}
+
\frac{1}{N}
\sum_{i=1}^{N}
w_i
\left(
y_i-\mu_i
\right)^2 ,
\end{equation}
where the weights $w_i$ are obtained from an inverse Kernel Density Estimation (KDE) of the redshift distribution,
\begin{equation}
w_i \propto \frac{1}{\hat p(z_i)}.
\end{equation}
This weighting scheme compensates for the strong concentration of Pantheon+ supernovae at low redshift and encourages the network to learn an accurate mean reconstruction across the entire redshift range. The contribution of the reconstruction term is gradually annealed to zero during the warmup period, providing a smooth transition from deterministic regression to fully probabilistic optimization. After the warmup phase, training proceeds using only the NLL objective defined above.
The predictive variance $\sigma_{\mathrm{pred}}(z)$ should not be interpreted as a direct estimate of a specific physical noise source. Instead, it represents a learned heteroscedastic predictive variance that complements the experimental covariance structure during training. In this way, the network is able to construct a probabilistic surrogate model whose predictive uncertainty remains well-defined even at redshifts where no direct observational covariance information is available.
Epistemic uncertainty is quantified independently through Monte Carlo Concrete Dropout sampling. Following training, dropout remains active at inference time and $N_{\mathrm{MC}}=1000$ stochastic forward passes are performed to generate samples from the approximate posterior predictive distribution. The variance of these realizations provides an estimate of the model uncertainty associated with limited data coverage and uncertainty in the network parameters. The final predictive uncertainty is obtained by combining this epistemic component with the learned heteroscedastic predictive variance.
Training is performed using stochastic gradient descent with momentum and an early-stopping criterion based on the validation loss. Specifically, training is halted if the validation loss fails to improve beyond the mean of the previous 30 evaluations for 3000 consecutive epochs. To assess convergence and calibration, we monitor both the NLL and the RMSE on the validation sample throughout training.

Explicit coverage assessment is central to the method. Although Bayesian and neural-network techniques have previously been used in supernova and cosmological applications \citep{moller20,mandel09,mancarella22,dialektopoulos23}, a probabilistic output does not itself ensure calibrated regression uncertainties. Following \citet{kfe18}, we compare predictive probabilities with empirical frequencies and measure the area between the reliability curve and the identity. We also track sharpness, defined here as mean predictive variance, to avoid evaluating coverage in isolation.

\begin{figure}
\centering
\includegraphics[width=\columnwidth]{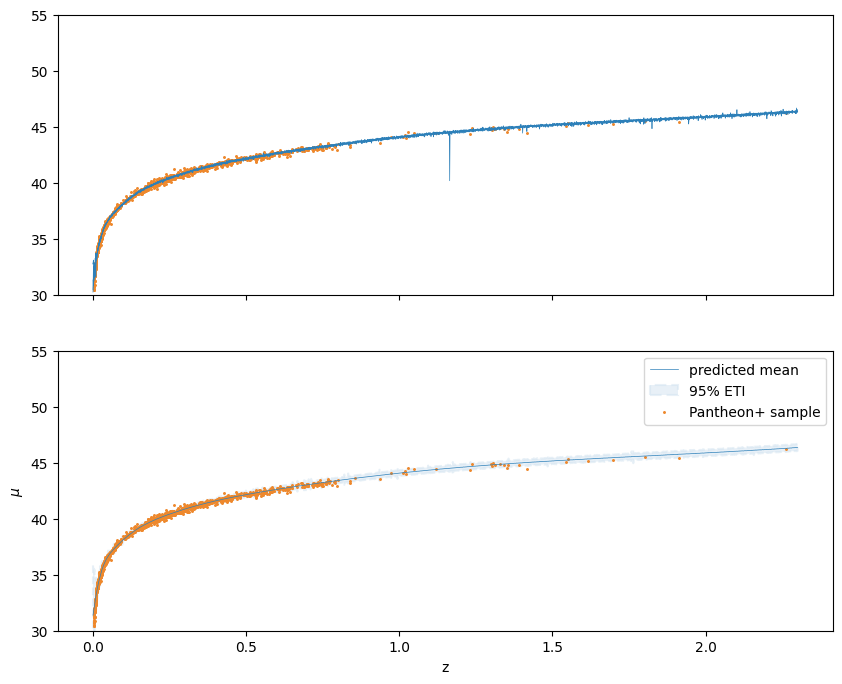}
\caption{BNN reconstruction of the Pantheon+ distance-modulus relation. Upper panel: stochastic predictive realizations. Lower panel: predictive mean and calibrated 95\% equal-tailed interval, with the supernova measurements in orange.}
\label{fig:bnn-reconstruction}
\end{figure}

A global uncertainty rescaling is fitted only on the calibration subset with the Uncertainty Toolbox \citep{chung21}; its performance is then assessed on the test subset. Metrics averaged over five independent trainings give a miscalibration-area reduction from 0.09 to 0.02 and a mean-variance reduction from 0.264 to 0.126. A Gaussian Q--Q diagnostic shows close agreement over most of the residual distribution, with mild tail departures. Appendix~\ref{app:bnn} retains the training curves, Q--Q plot, before/after reliability diagrams, and the procedure used to suppress non-reproducible Monte Carlo variance excursions. The final reconstructed relation is shown in Fig.~\ref{fig:bnn-reconstruction}.

\section{GRB standardization and Hubble diagram}\label{sec:calibration}
\subsection{Calibration likelihoods}
The calibrated BNN predicts distance moduli at the redshifts of the low-redshift GRBs. We convert them to $E_{\rm iso}$ using Eq.~\ref{eqEiso}, propagating both the fluence and distance uncertainties. Define
\begin{equation}\label{eq:xy}
x_i=\log_{10}\left(\frac{E_{{\rm p,i}}}{300\,{\rm keV}}\right),
\qquad y_i=\log_{10}\left(\frac{E_{{\rm iso},i}}{\rm erg}\right).
\end{equation}
For independent fluence and distance errors,
\begin{equation}
\sigma_{y,i}^2=\frac{1}{(\ln10)^2}
\left[\left(\frac{\sigma_{S,i}}{S_i}\right)^2+
4\left(\frac{\sigma_{d_L,i}}{d_{L,i}}\right)^2\right].
\end{equation}
We sample $(a,b,\sigma_{\rm int})$ with \texttt{emcee}, using both the D'Agostini and Reichart prescriptions \citep{dagostini05,reichart01}. With
\begin{equation}
v_i=\sigma_{\rm int}^2+\sigma_{y,i}^2+a^2\sigma_{x,i}^2,
\qquad r_i=y_i-b-a x_i,
\end{equation}
their expressions can be written compactly as
\begin{align}\label{eq:cal-likelihoods}
\mathcal L_{\rm D}&\propto\prod_i v_i^{-1/2}
\exp\left(-\frac{r_i^2}{2v_i}\right),\\
\mathcal L_{\rm R}&\propto(1+a^2)^{N/2}\mathcal L_{\rm D}.
\end{align}
The slope-dependent geometrical normalization distinguishes the Reichart prescription from a regression with intrinsic scatter assigned to the dependent variable. It changes the inferred slope; the two likelihoods should therefore be compared as distinct prescriptions, rather than treated as interchangeable normalizations. In the present parametrization, $\sigma_{\rm int}$ enters the residual variance $v_i$; the corresponding orthogonal scatter is $\sigma_{\rm int}/\sqrt{1+a^2}$.

\begin{figure}
\centering
\includegraphics[width=\columnwidth]{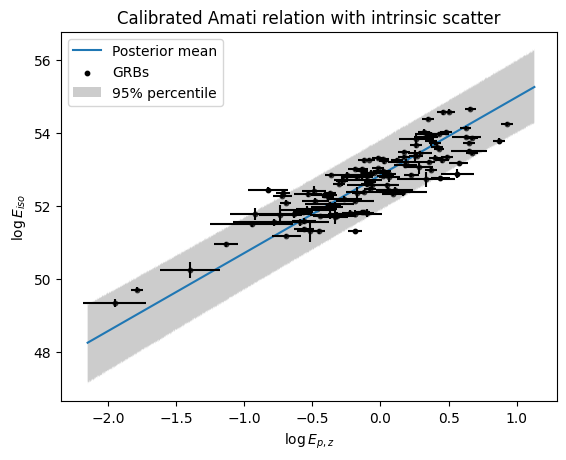}
\caption{Low-redshift Amati-relation calibration. The solid line is the posterior mean of the Reichart fit; the shaded region is the 95\% posterior predictive interval, including intrinsic scatter and parameter uncertainty. Error bars include uncertainties on both observables.}
\label{fig:fitamati}
\end{figure}

Table~\ref{tab:amati_params} gives the baseline BNN results, and Fig.~\ref{fig:fitamati} shows the Reichart fit used to construct the baseline Hubble diagram. For each likelihood prescription, the inferred parameters are consistent with earlier determinations obtained using comparable regression conventions. For example, \citet{huang25} reconstruct $H(z)$ with an artificial neural network, while \citet{xie25} use Gaussian processes with Pantheon+. The latter study obtains D'Agostini and Reichart slopes of $1.59^{+0.10}_{-0.10}$ and $2.026^{+0.083}_{-0.093}$, respectively, for the J221 sample at $z<1.4$. The full comparison values are retained in Appendix~\ref{app:calibration}.

\begin{table}
    \caption{Best fit parameters for $E_{\rm p,i}$ - $E_{\rm iso}$ correlation from our BNN analysis, using both the D'Agostini and Reichart likelihoods.}
    \label{tab:amati_params}
    \renewcommand{\arraystretch}{1.15}
    \centering
    \begin{tabular}{lccc}
        \hline
        \hline
        Method & \( b \) & \( a \) & \( \sigma_{int} \) \\
        \hline
        \noalign{\vskip 2mm}
        D'Agostini & $52.83_{-0.08}^{+0.08}$  & $1.67_{-0.17}^{+0.17}$ & $0.436$ \\
        Reichart   & $52.85_{-0.09}^{+0.09}$ & $2.13_{-0.20}^{+0.24}$ & $0.472$ \\
        \noalign{\vskip 2mm}
        \hline
        \hline
    \end{tabular}
\end{table}

\subsection{Distances and propagated uncertainties}\label{sec:grb-distances}
Assuming the calibrated relation remains applicable at higher redshift, the inferred energy gives
\begin{equation}\label{lumdist1}
d_L=\left[\frac{E_{\rm iso}(1+z)}{4\pi S_{\rm bolo}}\right]^{1/2},
\qquad
\mu=25+5\log_{10}\left(\frac{d_L}{\rm Mpc}\right),
\end{equation}
where $d_L$ from the first expression is in centimetres for cgs energies and fluences; conversion uses $1\,{\rm Mpc}=3.0857\times10^{24}$ cm. The uncertainty is
\begin{align}\label{eq:mu_unc}
\sigma_\mu^2&=\left(\frac52\right)^2\sigma_y^2+
\left(\frac{5}{2\ln10}\frac{\sigma_S}{S}\right)^2,\\
\sigma_y^2&=\sigma_{\rm int}^2+a^2\sigma_x^2+
\mathbf J\mathbf C_{ab}\mathbf J^T,\qquad \mathbf J=(x,1),\label{sigma_log_Eiso}
\end{align}
with $\sigma_x=\sigma_{E_{\rm p,i}}/(E_{\rm p,i}\ln10)$ and $\mathbf C_{ab}$ the slope--intercept covariance. The last term is $x^2\sigma_a^2+\sigma_b^2+2x\,\operatorname{Cov}(a,b)$. Thus parameter covariance, measurement uncertainty, and intrinsic dispersion all contribute; the intrinsic scatter currently dominates individual GRB errors.

\begin{figure}
\centering
\includegraphics[width=\columnwidth]{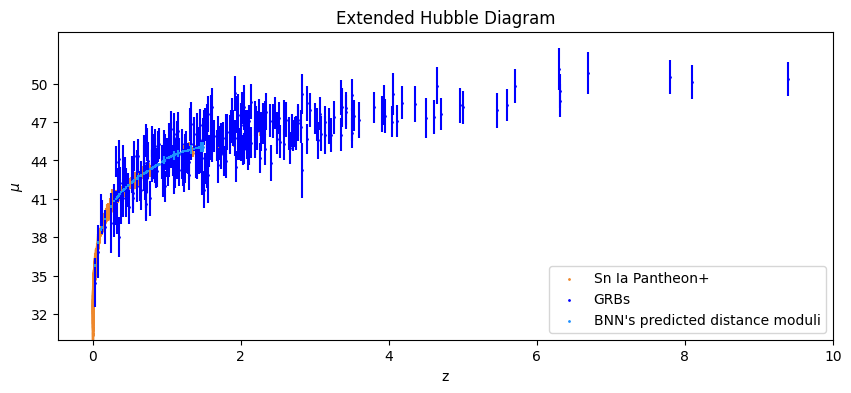}
\caption{Extended Hubble diagram. Blue points are GRB distance moduli, orange points are Pantheon+ SNIa, and light-blue points show the BNN predictions. Vertical bars represent 95\% equal-tailed intervals.}
\label{fig:extendedhubblediagram}
\end{figure}

The resulting diagram is shown in Fig.~\ref{fig:extendedhubblediagram}. Its distance scale is inherited from Pantheon+, so it is a high-redshift extension of the supernova ladder. An independently normalized GRB diagram would require an additive zero-point parameter degenerate with $H_0$. Moreover, anchoring to Pantheon+ induces statistical dependence between GRBs and SNIa. We retain the baseline cosmological analysis and explicitly assess this dependence using the controlled full-covariance test in Sect.~\ref{sec:sn-grb-dependence}. The complete calibrated diagram is provided as a machine-readable table
and is also reproduced in Appendix~\ref{app:data}.

\section{Cosmological inference}\label{sec:cosmology}
\subsection{Expansion histories and complementary probes}
We compare flat $\Lambda$CDM, non-flat $\Lambda$CDM, flat constant-$w$ dark energy, and the flat Chevallier--Polarski--Linder (CPL) parametrization \citep{cp01,linder03}. Radiation is neglected for the late-time distance calculation. With $\Omega_{\rm de}=1-\Omega_m-\Omega_k$,
\begin{align}\label{eq:ezfull}
E^2(z)&\equiv\frac{H^2(z)}{H_0^2}
=\Omega_m(1+z)^3+\Omega_k(1+z)^2+\Omega_{\rm de}f(z),\\
w(z)&=w_0+w_1\frac{z}{1+z},\label{cpleos}\\
f(z)&=(1+z)^{3(1+w_0+w_1)}
\exp\left[-\frac{3w_1z}{1+z}\right].\nonumber
\end{align}
Constant-$w$ models have $w_1=0$, and $\Lambda$CDM also fixes $w_0=-1$. Spatial curvature is free only in the non-flat $\Lambda$CDM case. We use $w_1$ for the CPL evolution parameter, also denoted $w_a$ in some diagnostics.

The radial comoving distance and luminosity distance are
\begin{equation}\label{distDC}
D_C=D_H\int_0^z\frac{d\zeta}{E(\zeta)},\qquad
D_L=(1+z)D_M,\qquad D_H=\frac{c}{H_0},
\end{equation}
where
\begin{equation}\label{eqdM}
D_M=\begin{cases}
\dfrac{D_H}{\sqrt{\Omega_k}}\sinh\!\left(\sqrt{\Omega_k}\,D_C/D_H\right),&\Omega_k>0,\\[2pt]
D_C,&\Omega_k=0,\\[2pt]
\dfrac{D_H}{\sqrt{|\Omega_k|}}\sin\!\left(\sqrt{|\Omega_k|}\,D_C/D_H\right),&\Omega_k<0.
\end{cases}
\end{equation}
BAO constrain combinations of $D_M$, $D_H=c/H(z)$, and $D_V=[D_M^2{cz/H(z)}]^{1/3}$ with the drag-epoch sound horizon $r_d$. We treat $r_d$ as free to avoid imposing an external sound-horizon calibration. The distance predictions are evaluated with Astropy.

The data combination comprises the calibrated GRBs, Pantheon+ \citep{brout22}, 34 observational Hubble measurements \citep{alm24,favale24,moresco22}, and DESI BAO \citep{adame25}. Cosmic chronometers estimate $H(z)=-(1+z)^{-1}dz/dt$ from differential ages of passively evolving galaxies \citep{moresco11}. The Hubble covariance embeds the released correlated 15-object block, with the remaining points treated as independent; the measurements are reproduced in Appendix~\ref{app:data}.

BAO measurements trace the sound-horizon imprint in galaxy clustering and the Lyman-$\alpha$ forest, complementing earlier low-redshift determinations \citep{beutler11,ross15,alam17}. We use the released DESI DR2 distance-ratio vector and its full covariance, including correlations between observables and redshift bins. These are predictions for $D_H/r_d$, $D_M/r_d$, or $D_V/r_d$, as appropriate to each released measurement.

For completeness, the drag-epoch ruler is
\begin{equation}
r_d=\int_{z_d}^{\infty}\frac{c_s(z)}{H(z)}\,dz,
\end{equation}
where $c_s$ is the sound speed of the baryon--photon plasma and $z_d$ is the drag-epoch redshift. This definition specifies the ruler entering the observations; $r_d$ is nevertheless a free parameter in our inference. The angular-diameter distance satisfies $D_A=D_M/(1+z)$. Transverse BAO constrain $D_M/r_d$, radial BAO constrain $D_H/r_d$, and an isotropic combination constrains $D_V/r_d$. They therefore complement the luminosity-distance measurements rather than provide another standard-candle normalization.

The DESI measurement vector combines tracers over different redshift intervals. These include the Lyman-$\alpha$ forest at $1.77<z<4.16$, luminous red galaxies at $0.4<z<0.8$, emission-line galaxies at $1.1<z<1.6$, and the combined red-/emission-line-galaxy sample at $0.8<z<1.1$. The bright-galaxy and quasar samples supply isotropic distance information in the adopted data vector. All theoretical entries are matched to the released observable definitions and covariance, rather than treating transverse and radial estimates as independent measurements. The shared sound-horizon scale is sampled jointly with the expansion-history parameters.

For every data block $X$, the baseline statistic has the Gaussian form
\begin{equation}\label{eq:joint-baseline}
\chi_X^2=(\mathbf D_X-\mathbf m_X)^T\mathbf C_X^{-1}
(\mathbf D_X-\mathbf m_X),\qquad
\chi^2_{\rm tot}=\sum_X\chi_X^2,
\end{equation}
where $X=\{{\rm GRB,SN,OHD,DESI}\}$. Pantheon+ retains its full covariance. The separate SN and GRB blocks are a baseline approximation whose impact is quantified in Sect.~\ref{sec:sn-grb-dependence}; they should not be interpreted as statistically independent observations of the absolute distance scale. We sample each model with \texttt{emcee}, using the uniform priors in Table~\ref{priors-cosmomPasq}.

\begin{table}
\caption{The priors used in our cosmological analysis. }
\resizebox{\columnwidth}{!}{
	\begin{tabular}{c|c c  c c c c }
		\hline \hline
		Model  & $\Omega_m$ & $\Omega_\Lambda$ & h & $r_d$ & $w_0$ & $w_1$  \\
		\hline
        $\Lambda$CDM & $\{0.01, 1.00\}$ & & $\{0.2,1\}$ & $\{120, 150\}$ &  & \\
        \hline
		non-flat-$\Lambda$CDM & $\{0.01, 1.00\}$ & $\{0.001 , 1.00\}$& $\{0.2,1\}$ & $\{ 120, 150\}$ &  & \\
		\hline
	    wCDM & $\{0.01 , 1.00\}$ & &  $\{0.2,1\}$ & $\{120, 150\}$ & $\{-3,  1\}$ & \\
		\hline
		 CPL & $\{0.01, 1.00\}$ & & $\{0.2,1\}$ & $\{120, 150\}$ & $\{-3, 1\}$ &$\{ -3 , 2\}$ \\
		\hline \hline
	\end{tabular}} \label{priors-cosmomPasq}
\end{table}

\subsection{Model comparison and interpretation}
\begin{figure*}
\centering
\begin{minipage}{0.48\textwidth}\centering
\includegraphics[width=\linewidth]{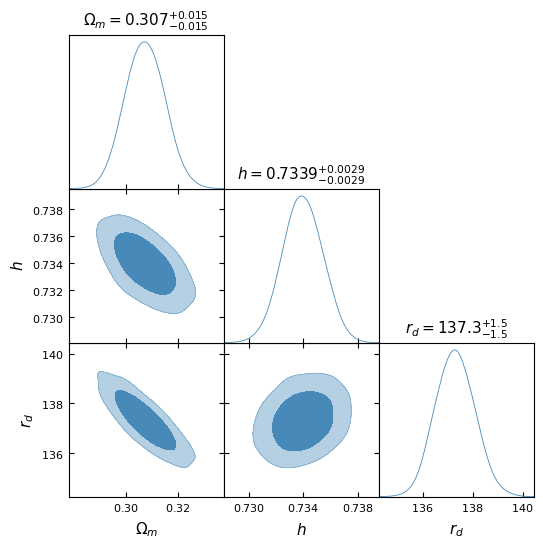}\\(a) Flat $\Lambda$CDM
\end{minipage}\hfill
\begin{minipage}{0.48\textwidth}\centering
\includegraphics[width=\linewidth]{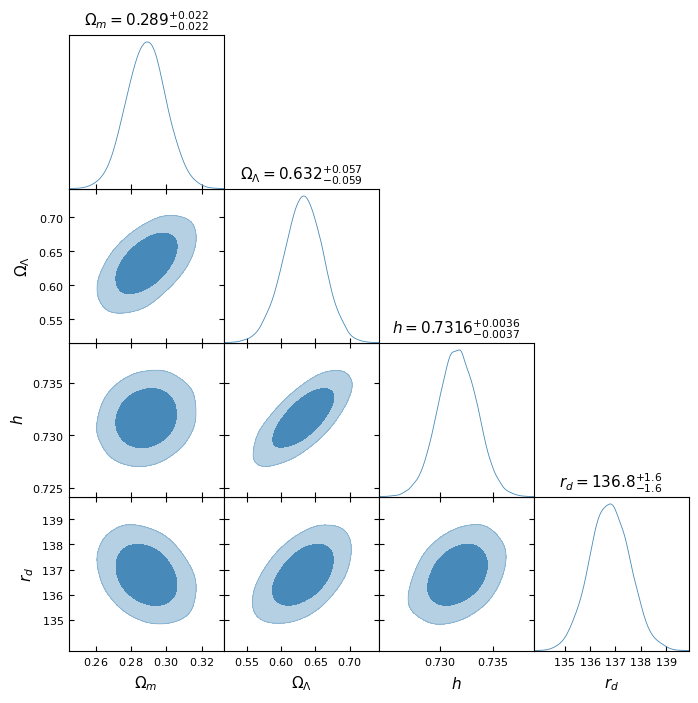}\\(b) Non-flat $\Lambda$CDM
\end{minipage}
\par\medskip
\begin{minipage}{0.48\textwidth}\centering
\includegraphics[width=\linewidth]{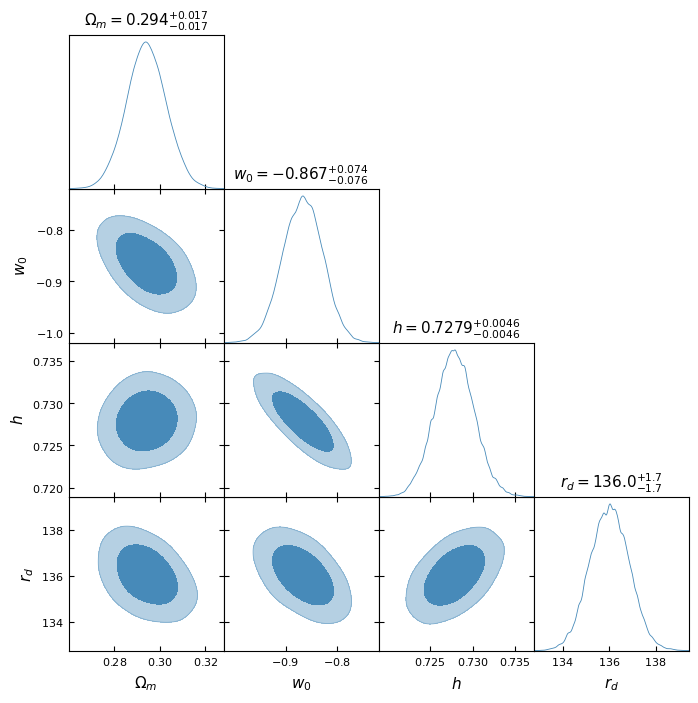}\\(c) Flat constant-$w$ model
\end{minipage}\hfill
\begin{minipage}{0.48\textwidth}\centering
\includegraphics[width=\linewidth]{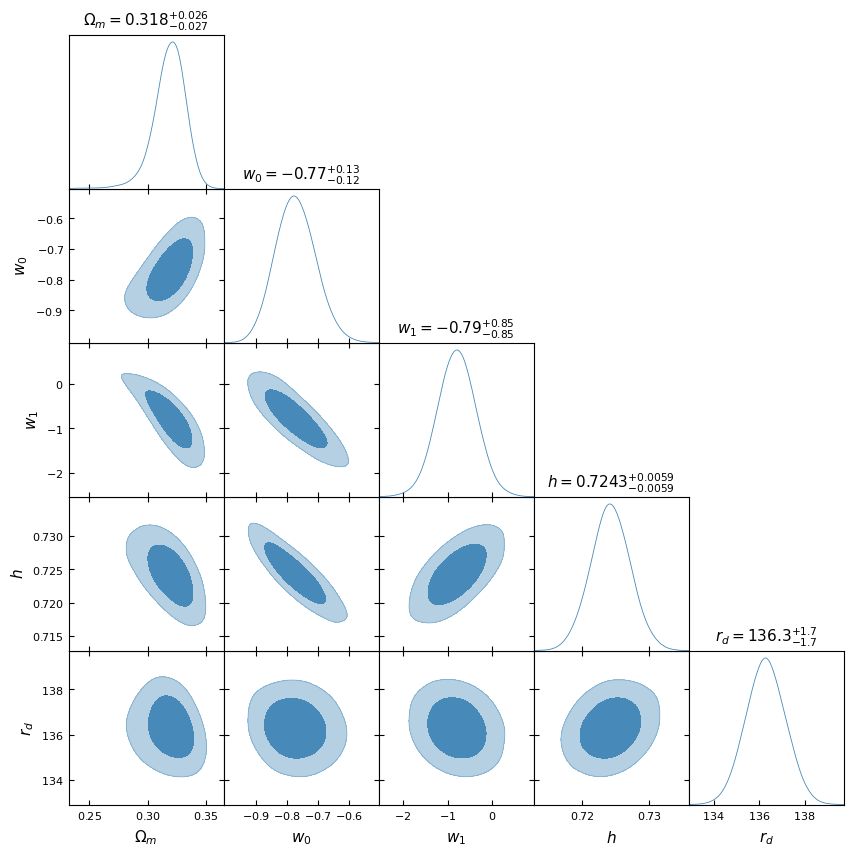}\\(d) Flat CPL model
\end{minipage}
\caption{One-dimensional marginalized posteriors and two-dimensional 68\% and 95\% credible regions for the four cosmological fits. Panels retain the original parameter estimates and contours. The non-flat model includes $\Omega_\Lambda$; the constant-$w$ and CPL models introduce the corresponding dark-energy parameters.}
\label{fig:cosmo-posteriors}
\end{figure*}

The marginalized posteriors and two-dimensional $1\sigma$ and $2\sigma$ contours are shown in Fig.~\ref{fig:cosmo-posteriors}. Within the adopted fit, the joint CPL posterior places the cosmological-constant point at an approximately \(2\sigma\) displacement from the posterior maximum. Such a displacement is conditional on the data combination, priors, and calibration, and is not by itself a model-selection result.

We compare the cosmological fits with the Akaike and Bayesian information criteria \citep{akaike92,whr12},
\begin{equation}\label{aic}
{\rm AIC}=\chi^2_{\min}+2K,\qquad
{\rm BIC}=\chi^2_{\min}+K\ln N,
\end{equation}
The AIC and BIC values are calculated from the corresponding
minimum chi-square values.
\begin{table}
\centering
\caption{Information criteria for the observed-data fits. Differences are recomputed from the displayed AIC and BIC values relative to their respective minima; $N=1701_{\rm SN}+264_{\rm GRB}+34_{\rm OHD}+12_{\rm BAO}=2011$.}
\label{aicbic}
\setlength{\tabcolsep}{3.5pt}
\begin{tabular}{lrrrrr}
\toprule
Model & $K$ & AIC & $\Delta$AIC & BIC & $\Delta$BIC\\
\midrule
Flat $\Lambda$CDM & 3 & 2011.24 & 9.71 & 2028.06 & 2.39\\
Non-flat $\Lambda$CDM & 4 & 2010.42 & 8.89 & 2032.84 & 7.17\\
Flat $w$CDM & 4 & 2003.25 & 1.72 & 2025.67 & 0.00\\
Flat CPL & 5 & 2001.53 & 0.00 & 2029.56 & 3.89\\
\bottomrule
\end{tabular}
\end{table}

AIC selects CPL, whereas BIC selects the simpler constant-$w$ model. Consequently, the present comparison supports improved phenomenological fits from dark-energy extensions within this framework, but does not establish evolution of the equation of state. In particular, the BIC evidence against flat \(\Lambda\)CDM is only weak to positive. This distinction between an extension with $w\ne-1$ and evidence for $w_1\ne0$ is essential when interpreting the high-redshift information carried by the GRBs.

\FloatBarrier
\section{Robustness of the standardization}\label{sec:robustness}
\subsection{Local-regression and likelihood comparisons}\label{sec:local-regression}
We compare the BNN reconstruction with local regression on Pantheon+, following the approach of \citet{demianski21} but propagating the full supernova covariance. At each GRB redshift, a weighted linear fit to neighbouring SNIa gives the reconstructed modulus. Writing the resulting operator as $\widehat{\boldsymbol\mu}=\mathbf A\boldsymbol\mu_{\rm SN}$ makes the propagated covariance explicit:
\begin{equation}\label{eq:lr-cov}
\mathbf C_\mu=\mathbf A\mathbf C_{\rm SN}\mathbf A^T,
\qquad
\mathbf C_y=\left(\frac25\right)^2\mathbf C_\mu+
\operatorname{diag}\left[\left(\frac{\sigma_S}{S\ln10}\right)^2\right].
\end{equation}
Appendix~\ref{app:calibration} specifies the neighbourhood weights and full-covariance likelihood. The full-covariance Reichart-type version retains the slope-dependent normalization but replaces the independent residual distribution by a multivariate Gaussian. It is a pragmatic diagnostic prescription, rather than a unique extension of the independent-point Reichart derivation. It does not replace the baseline BNN Hubble diagram.

The local-regression D'Agostini-type fit gives $a=1.66\pm0.09$, $b=52.80\pm0.04$, and $\sigma_{\rm int}=0.43\pm0.03$. The Reichart-type diagnostic gives $a=2.13\pm0.11$, $b=52.84\pm0.05$, and $\sigma_{\rm int}=0.47\pm0.04$. Thus the steeper Reichart slope is reproduced with a different distance reconstruction. A local-regression test using approximately 3000 mock SNIa extracted from the BNN gives $a=1.67\pm0.08$, $b=52.60\pm0.06$, and $\sigma_{\rm int}=0.43\pm0.03$. The slope and scatter are stable; this comparison does not require identical intercepts. The supplied comparison table, with all quoted asymmetric errors, is reproduced in Appendix~\ref{app:calibration}.

\subsection{Impact of the SN--GRB statistical dependence}
\label{sec:sn-grb-dependence}
Since the GRB calibration ultimately relies on the Pantheon+ distance
scale, the SNIa and GRB contributions to the cosmological likelihood
are not, strictly speaking, statistically independent. We quantified the size
of this effect through an explicit controlled diagnostic based on the
local-regression, full-covariance calibration framework and the
Reichart likelihood.
Let $C_{\rm SN}$ denote the Pantheon+ covariance matrix and let $A$ be
the linear local-regression operator mapping the SN distance moduli
onto the GRB calibration distances. The covariance propagated to the
calibration sample is therefore
\begin{equation}
C_{\mu,{\rm cal}} = A\,C_{\rm SN}\,A^{T}.
\end{equation}
The response of the Reichart calibration parameters to perturbations
of the reconstructed calibration distances was evaluated through the
implicit likelihood Jacobian. Propagating this response to the 264
GRB distance moduli yields an explicit SN--GRB cross-covariance block
$C_{\rm SN,GRB}$.
As a consistency check, the corresponding SN-induced component of the
GRB covariance,
\begin{equation}
C_{\rm shared}
=
C_{\rm SN,GRB}^{T}
C_{\rm SN}^{-1}
C_{\rm SN,GRB},
\end{equation}
is positive semidefinite within numerical precision. When combined
with the independently reconstructed Reichart GRB covariance, the
conditional GRB covariance is given by the Schur complement
\begin{equation}
S_{\rm GRB}
=
C_{\rm GRB}
-
C_{\rm SN,GRB}^{T}
C_{\rm SN}^{-1}
C_{\rm SN,GRB}.
\end{equation}
The latter remains positive definite, with a minimum eigenvalue
$\lambda_{\min}=1.44875$. The largest-eigenvalue ratio
\begin{equation}
\frac{\lambda_{\max}(C_{\rm shared})}
{\lambda_{\max}(C_{\rm GRB})}
\simeq 0.0534
\end{equation}
shows that the shared component is non-zero but subdominant. This
quantity is a global spectral diagnostic and should not be interpreted
as the fraction of the uncertainty of each individual GRB that is
inherited from the SN sample.
We then compared the standard ``block'' likelihood, in which the SN
and GRB terms are treated as independent, with the full joint
Gaussian likelihood including $C_{\rm SN,GRB}$. Over 2800 samples
drawn from the converged cosmological posterior, the difference
$\Delta\chi^2=\chi^2_{\rm joint}-\chi^2_{\rm block}$ has median
$-1.252$ and standard deviation $0.217$. Thus, the cross-covariance
produces a measurable modification of the likelihood, although its
impact on the cosmological constraints is small. The largest shift of
a posterior median is $0.040\,\sigma$, while the largest relative
change of a 68\% posterior width is $2.71\%$ (for $w_0$). Here the
normalized median shift is defined as
\begin{equation}
\Delta_{\rm norm}(\theta)
=
\frac{
\theta_{\rm joint}^{\rm med}-\theta_{\rm block}^{\rm med}
}{
\sqrt{\sigma_{\rm joint}^{2}+\sigma_{\rm block}^{2}}
},
\qquad
\sigma=\frac{q_{84}-q_{16}}{2},
\end{equation}
where $q_{16}$ and $q_{84}$ denote the 16th and 84th posterior
quantiles, respectively. The corresponding normalized shifts for
$\Omega_{\rm m}$, $w_1$, and $h$ are approximately
$0.031\,\sigma$, $0.0023\,\sigma$, and $0.0011\,\sigma$,
respectively.
We therefore conclude that the SN--GRB statistical dependence should
be acknowledged at the covariance level, but within this explicit
full-covariance diagnostic it does not significantly affect the
cosmological conclusions.
\subsection{Leave-three-highest-redshift-out calibration test}
\label{sec:highz-leaveout}
We also tested the sensitivity of the GRB calibration to the inclusion
of the three highest-redshift Pantheon+ SNe, at
$z=1.80119$, $1.91165$, and $2.26137$. Since the original BNN
training products needed to retrain the network after removing these
objects are unavailable, we performed a controlled leave-three-out
test within the explicit local-regression, full-covariance calibration
framework.
The local regression uses a fraction $\alpha=0.02$ of the Pantheon+
sample. For the complete sample this corresponds to 34 neighbouring
SNe. Changing the sample size can also change the integer neighbourhood
selected by the fractional rule, mixing removal of the high-$z$ SNe with
a discrete change of the regression algorithm. We therefore adopted as our
primary diagnostic a fixed-$n=34$ leave-three-out reconstruction,
while retaining the nominal $\alpha=0.02$ reconstruction as an
additional algorithmic robustness check.
In the fixed-$n=34$ test, only 20 of the 123 calibration
reconstructions are modified, all in the interval
$1.293\leq z_{\rm GRB}\leq1.489$, exactly as predicted from the
neighbour-selection analysis. The largest change of an individual
calibration distance modulus is
\begin{equation}
\max |\Delta\mu_{\rm cal}| = 0.02338~{\rm mag},
\end{equation}
and the largest standardized change is $1.76\,\sigma$. The latter
corresponds to a very small absolute shift
($|\Delta\mu_{\rm cal}|\simeq 0.00193$ mag), while the largest
absolute shift, occurring at the upper calibration boundary, is
approximately a $1\,\sigma$ effect.
Refitting the Reichart relation after removing the three SNe yields 
\begin{align}
\Delta a &= 1.04\times10^{-4},\\
\Delta b &= 4.16\times10^{-4},\\
\Delta\sigma_{\rm int} &= 8.84\times10^{-5}.
\end{align}
The shifts in $a$ and $b$ are below $0.01\,\sigma$ relative to their
posterior uncertainties.
The corresponding change in the 264-GRB Hubble diagram is extremely
small,
\begin{equation}
\max |\Delta\mu_{\rm GRB}|
=
1.31\times10^{-3}~{\rm mag},
\end{equation}
with a maximum standardized displacement of only
$1.0\times10^{-3}\,\sigma$. The predictive GRB covariance is equally
stable: the relative Frobenius-norm variation is
\begin{equation}
\frac{\|\Delta C_{\rm GRB}\|_F}
     {\|C_{\rm GRB}\|_F}
=
3.19\times10^{-4},
\end{equation}
the maximum change of an individual GRB uncertainty is $0.0183\%$,
and the covariance remains positive definite.
Finally, propagating the leave-three-out calibration to the
cosmological posterior produces no resolved shift of the posterior
medians at the resolution of our importance sample. The largest
relative change of a 68\% posterior width is only $0.129\%$.
The importance reweighting is well conditioned, with
${\rm ESS}/N\simeq0.88$.
These results show that, within this controlled leave-three-out
diagnostic, the three highest-redshift Pantheon+ SNe do not drive the
GRB calibration or the inferred cosmological constraints. As for the
cross-covariance test above, this analysis is intended as an explicit
sensitivity diagnostic rather than as an exact retraining of the
original BNN.
\subsection{Parametric test for redshift evolution}
\label{sec:redshift-evolution}
\begin{figure}
    \centering
    \includegraphics[width=\columnwidth]{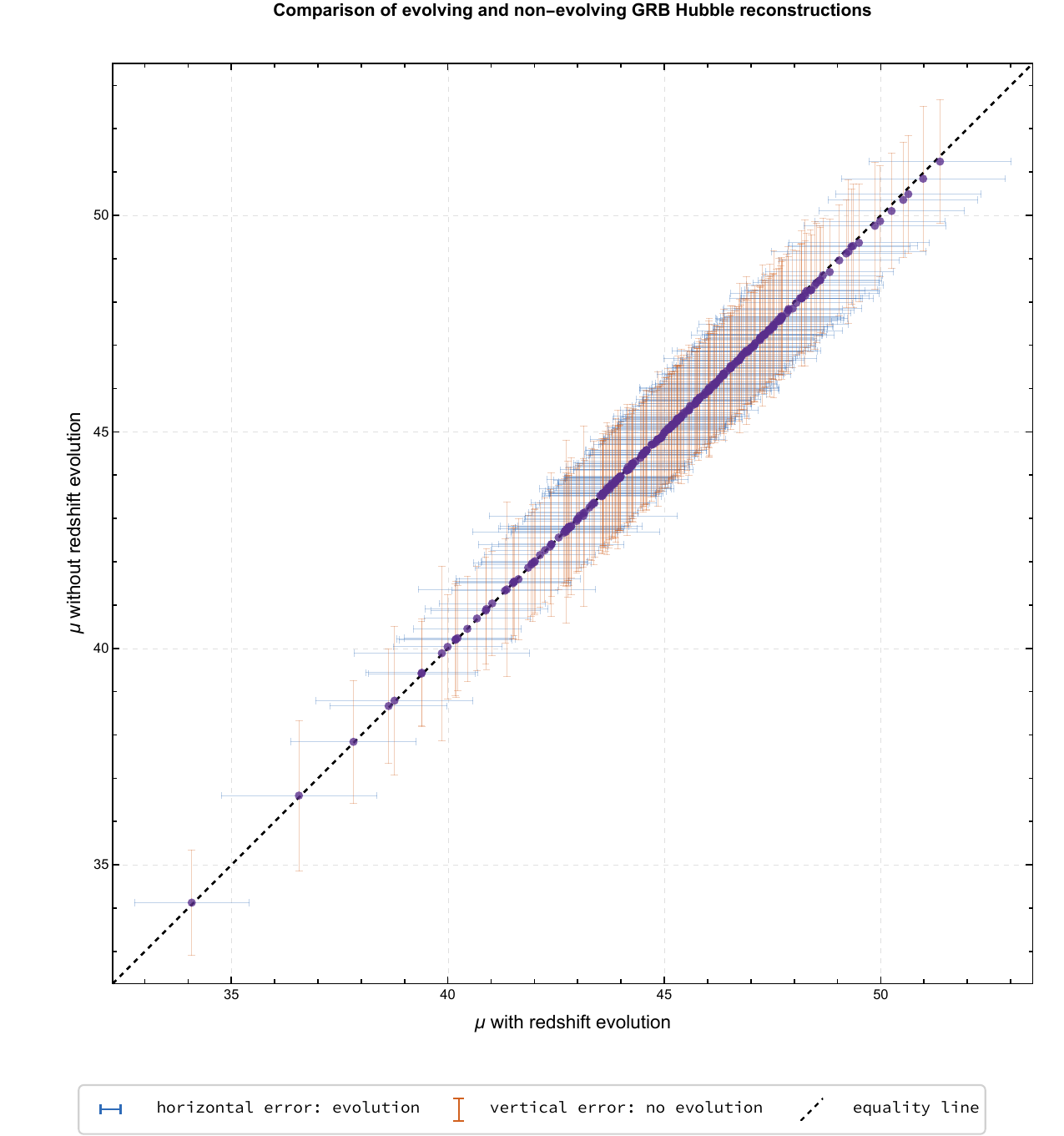}
    \caption{One-to-one comparison between the GRB distance moduli reconstructed with and without the redshift-evolution term. Horizontal error bars refer to the evolving reconstruction, while vertical error bars refer to the non-evolving case. The dashed line denotes $\mu_{\rm no\text{-}evo}=\mu_{\rm evo}$. Identical ranges are adopted on both axes.}
    \label{fig:mu-evolution-comparison}
\end{figure}

To assess the possible redshift dependence of the $E_{\rm p,i}$--$E_{\rm iso}$ correlation, we extended the calibration model to
\begin{equation}
\log_{10}\!\left(\frac{E_{\rm iso}}{\mathrm{erg}}\right)
=
a\,\log_{10}\!\left(\frac{E_{\rm p,i}}{300\,\mathrm{keV}}\right)
+b
+\gamma \log_{10}(1+z),
\end{equation}
while retaining the same full-covariance Reichart-type likelihood adopted in the non-evolving analysis. The additional parameter $\gamma$ was sampled jointly with $a$, $b$, and $\sigma_{\rm int}$, and the four independent MCMC chains reached satisfactory convergence, with split-$\widehat{R}<1.01$ for all parameters. The marginalized posterior constraints are summarized in Table~\ref{tab:redshift-evolution-posterior}.
\begin{table}
\centering
\caption{Marginalized posterior constraints on the parameters of the redshift-dependent $E_{\rm p,i}$--$E_{\rm iso}$ relation. The quoted values are posterior medians with equal-tailed 68\% and 95\% credible intervals.}
\label{tab:redshift-evolution-posterior}
\begin{tabular}{lcc}
\toprule
Parameter & 68\% credible interval & 95\% credible interval \\
\midrule
$a$
& $2.12^{+0.136}_{-0.123}$
& $2.12^{+0.283}_{-0.233}$ \\
$b$
& $52.8^{+0.167}_{-0.157}$
& $52.8^{+0.340}_{-0.307}$ \\
$\gamma$
& $0.0841^{+0.564}_{-0.599}$
& $0.0841^{+1.10}_{-1.22}$ \\
$\sigma_{\rm int}$
& $0.474^{+0.0460}_{-0.0392}$
& $0.474^{+0.0982}_{-0.0719}$ \\
\bottomrule
\end{tabular}
\end{table}
The posterior medians of $a$, $b$, and $\sigma_{\rm int}$ remain fully compatible with those obtained in the non-evolving analysis, indicating that the inclusion of the additional redshift-dependent term does not materially alter the calibration. The parameter $\gamma$, instead, is only weakly constrained and its credible intervals extend broadly across zero. The marginalized constraint,
\begin{equation}
\gamma = 0.084^{+0.564}_{-0.599}
\qquad (68\%~\mathrm{credible~interval}),
\end{equation}
is fully consistent with $\gamma=0$; the corresponding 95\% credible interval is
$\gamma = 0.084^{+1.098}_{-1.219}$. The posterior probabilities
$P(\gamma>0)=0.558$ and $P(\gamma<0)=0.442$ further show that the data do not provide a meaningful preference for either sign of the evolution. Although the deterministic best fit is obtained at $\gamma_{\rm BF}\simeq 0.211$, the improvement in likelihood with respect to the non-evolving model is negligible:
$2\Delta\ln\mathcal{L}=0.129$, corresponding to $p=0.719$ in the asymptotic likelihood-ratio test. Model-selection criteria lead to the same conclusion, with
$\Delta\mathrm{AIC}=+1.87$ and $\Delta\mathrm{BIC}=+4.68$, where the differences are defined as the evolving-model value minus the non-evolving-model value. Therefore, the introduction of the redshift-dependent term is not statistically warranted by the present calibration sample. Within this full-covariance local-regression analysis, the non-evolving parametrization is retained as the reference case, while the evolving model is used as a robustness test. This diagnostic comparison does not replace the BNN-based GRB Hubble diagram adopted for the cosmological constraints.
The practical impact of the evolution term on the reconstructed GRB Hubble diagram was further assessed through the one-to-one comparison shown in Fig.~\ref{fig:mu-evolution-comparison}, in the $\mu_{\rm evo}$--$\mu_{\rm no\text{-}evo}$ plane. Horizontal and vertical error bars represent, respectively, the uncertainties associated with the evolving and non-evolving reconstructions, while the line $\mu_{\rm no\text{-}evo}=\mu_{\rm evo}$ marks perfect agreement. Identical ranges were adopted on both axes in order to avoid any artificial visual enhancement of deviations from the identity relation. The points lie extremely close to the diagonal. Defining $\Delta\mu=\mu_{\rm no\text{-}evo}-\mu_{\rm evo}$, we find
\begin{eqnarray}
&&\langle\Delta\mu\rangle=-0.0366,\qquad\nonumber \\
&&\mathrm{median}(\Delta\mu)=-0.0330,\qquad \nonumber\\
&&\mathrm{rms}(\Delta\mu)=0.0533~\mathrm{mag}\nonumber,
\end{eqnarray}
with values ranging from $-0.1610$ to $0.0433$ mag. The negative mean indicates that the non-evolving reconstruction yields distance moduli that are, on average, only about $0.04$ mag smaller than those obtained when the evolution term is included. This displacement is negligible compared with the individual distance-modulus uncertainties. Indeed, the maximum absolute normalized difference,
\begin{equation}
\max\!\left[
\frac{|\Delta\mu|}
{\sqrt{\sigma_{\mu,\rm evo}^{2}+\sigma_{\mu,\rm no\text{-}evo}^{2}}}
\right]
=0.0738,
\end{equation}
is far below unity. Moreover, the two sets of reconstructed distance moduli are almost perfectly correlated, with a Pearson coefficient $r=0.999963$. These results demonstrate that allowing for redshift evolution produces only a very small, highly coherent shift of the Hubble diagram and no statistically relevant object-by-object discrepancy. The normalized-difference statistic should be regarded as descriptive, since the two reconstructions are derived from the same GRB sample and are therefore correlated; nevertheless, together with the posterior constraint on $\gamma$ and the model-selection results, it confirms the robustness of the non-evolving calibration. Moreover, although this test has been performed within the local-regression Reichart full-covariance
framework, it probes the same GRB calibration sample and the same
\(E_{\rm p,i}\)--\(E_{\rm iso}\) standardization assumption used in the BNN analysis.
\subsection{Truncation-corrected marginal evolution}\label{sec:ep-test}
The parameter $\gamma$ measures evolution of the relation at fixed $E_{\rm p,i}$, not evolution of the marginal energy distributions along it. We therefore also apply the Efron--Petrosian (EP) modified-rank statistic \citep{efron92,efron99,dainotti22}. Each object's rank is evaluated only within an associated set that could have been observed under the adopted truncation. Since exact detector-specific selection functions are unavailable, we use a single effective bolometric-fluence threshold as a controlled approximation.

The 10th-percentile baseline threshold is $S_{{\rm bolo},\lim}=2.26269\times10^{-6}$ erg cm$^{-2}$, retaining 238 of the 264 bursts. It defines $E_{{\rm iso},\lim}(z)=4\pi d_L^2S_{{\rm bolo},\lim}/(1+z)$. Removing trial evolution as $E'_{\rm iso}=E_{\rm iso}/(1+z)^k$ gives $\tau_{\rm EP}(0)=-3.1866$ and
\begin{equation}\label{eq:kiso-ep-main}
k_{\rm iso}=1.714^{+0.717}_{-0.555},\qquad
1.159<k_{\rm iso}<2.431,
\end{equation}
where $\tau_{\rm EP}=\pm1$ defines the conventional interval. Appendix~\ref{app:ep-details} retains the associated-set definition, threshold and fiducial-cosmology scans, and Monte Carlo diagnostics.

\begin{table}
\centering
\caption{Summary of the Efron--Petrosian robustness tests. Detailed definitions and numerical results are reported in Appendix~\ref{app:ep-details}.}
\label{tab:ep-robustness-main}
\small
\resizebox{\columnwidth}{!}{%
\begin{tabular}{lll}
\toprule
Diagnostic & Result for $k_{\rm iso}$ & Interpretation \\
\midrule
Baseline EP & $1.714^{+0.717}_{-0.555}$ & Primary interval \\
Fluence cuts 5--10\% & $1.714$--$1.981$ & Stable central values \\
$0.20\leq\Omega_m\leq0.40$ & $1.633$--$1.848$ & Weak cosmology dependence \\
Conditional MC & $1.735^{+0.115}_{-0.096}$ & Fluence-error sensitivity \\
Variable-membership MC & $1.769^{+0.170}_{-0.162}$ & Threshold sensitivity \\
\bottomrule
\end{tabular}%
}
\end{table}

Positive marginal energy evolution is compatible with $\gamma\simeq0$. If $E_{\rm p,i}$ evolves as $(1+z)^{k_p}$, then
\begin{equation}\label{eq:gamma-kiso-kp}
\gamma=k_{\rm iso}-a k_p.
\end{equation}
Combining the joint $(a,\gamma)$ posterior with the EP uncertainty gives the indirect estimate $k_p\simeq0.790^{+0.404}_{-0.385}$, with $P(k_p>0)\simeq0.98$. This suggests evolution predominantly along the relation, but is not a direct EP peak-energy measurement: bandpass and peak-energy selection have not been modelled. Threshold scans give central $k_{\rm iso}$ values from 1.714 to 1.981; changing $\Omega_m$ from 0.20 to 0.40 gives 1.633--1.848. Separate fixed-sample and variable-membership Monte Carlo experiments confirm stability within the adopted approximation. These checks do not amount to an end-to-end correction of the heterogeneous catalogue's instrumental and redshift-selection effects.

\section{Perspectives for THESEUS}\label{sec:theseus}
Future missions may provide larger and more homogeneous samples of high-redshift GRBs with measured prompt spectra and redshifts.
THESEUS (Transient High-Energy Sky and Early Universe Surveyor) is a proposed space-mission concept designed to exploit GRBs as probes of the early Universe and as electromagnetic counterparts of multimessenger transients. Its broad-band prompt-emission coverage and rapid redshift characterization motivate our consideration of samples extending to \(z\gtrsim3\), potentially approaching \(z\simeq10\). We assess the possible cosmological constraining power of such a sample using idealized, already-standardized distance-modulus catalogues; we do not simulate the mission selection process.

Each mock contains $\{z_i,\mu_i,\sigma_{\mu,i}\}$, with
\begin{align}\label{eq:mock-observable}
\mu_i^{\rm mock}&=\mu_{\rm th}(z_i;\theta_{\rm fid})+
\sigma_{\mu,i}\epsilon_i,\qquad\epsilon_i\sim\mathcal N(0,1),\\
\sigma_{\mu,i}^2&=\sigma_{\rm sys}^2+
d_m^2\left(\frac{1+z_i}{1+z_{\max}}\right)^2,\nonumber
\end{align}
where $\sigma_{\rm sys}=0.25$ mag and $d_m=0.05$. This expression defines the assigned variance of each simulated object in these mocks; a correlated systematic covariance is not introduced. A free additive offset $\Delta\mu$ absorbs the distance zero point and effective $H_0$ normalization, and is profiled analytically. The flat-$\Lambda$CDM and flat-CPL fits thus compare expansion-history shape rather than an externally fixed absolute calibration. Appendix~\ref{app:forecast} gives the profiled likelihood, parameter counts, and complete numerical tables.

We first analyse 100 $\Lambda$CDM-injected catalogues. The mean improvement $\Delta\chi^2=\chi^2_{\Lambda\rm CDM}-\chi^2_{\rm CPL}$ is 1.753. Only 1\% exceed the nominal two-parameter $2\sigma$ threshold, and none exceed $3\sigma$. Mean $\Delta{\rm AIC}=2.247$ and $\Delta{\rm BIC}=12.874$, defined as CPL minus $\Lambda$CDM, favour the generating $\Lambda$CDM model. The null experiment therefore shows no systematic spurious preference for dynamical dark energy under these assumptions.

\begin{table}
\centering
\caption{Null-test results from 100 $\Lambda$CDM mock catalogues. The statistic $\Delta\chi^2$ is defined as $\chi^2_{\Lambda {\rm CDM}}-\chi^2_{\rm CPL}$. Positive $\Delta {\rm AIC}$ and $\Delta {\rm BIC}$ values favour $\Lambda$CDM.}
\label{tab:lcdm_null_test}
\begin{tabular}{lc}
\hline
Quantity & Value \\
\hline
Number of mock catalogues & 100 \\
Mean $\Delta\chi^2$ & 1.753 \\
Median $\Delta\chi^2$ & 1.281 \\
Maximum $\Delta\chi^2$ & 7.444 \\
False positive rate above $1\sigma$ & 0.32 \\
False positive rate above $2\sigma$ & 0.01 \\
False positive rate above $3\sigma$ & 0.00 \\
Mean $\Delta {\rm AIC}$ & 2.247 \\
Mean $\Delta {\rm BIC}$ & 12.874 \\
\hline
\end{tabular}
\end{table}

The complementary experiment uses 50 independent catalogues of 1500 standardized GRBs generated from the CPL vector inferred in the supplied analysis. This is a conditional question: if that fiducial CPL cosmology and the adopted error model describe the future sample, how often would it be distinguishable from $\Lambda$CDM? The mean improvement is $\langle\Delta\chi^2\rangle=49.319$, all realizations exceed the nominal $3\sigma$ threshold, and both AIC and BIC favour CPL in every realization. The mean differences are $-45.319$ and $-34.692$, respectively. The evolution parameter remains the least stable recovered quantity: its realization-to-realization scatter is 3.006, and 6\% of fits meet its lower prior bound. Thus model discrimination and accurate recovery of every CPL parameter should not be conflated.

\begin{table}
\centering
\caption{Detection forecast from 50 CPL-injected mock catalogues. Negative $\Delta {\rm AIC}$ and $\Delta {\rm BIC}$ values favour CPL.}
\label{tab:cpl_detection_forecast}
\begin{tabular}{lc}
\hline
Quantity & Value \\
\hline
Number of requested mock catalogues & 50 \\
Number of valid fits & 50 \\
Mean $\Delta\chi^2$ & 49.319 \\
Median $\Delta\chi^2$ & 47.890 \\
Maximum $\Delta\chi^2$ & 90.539 \\
Detection rate above $1\sigma$ & 1.00 \\
Detection rate above $2\sigma$ & 1.00 \\
Detection rate above $3\sigma$ & 1.00 \\
Mean $\Delta {\rm AIC}$ & $-45.319$ \\
Median $\Delta {\rm AIC}$ & $-43.890$ \\
Fraction favouring CPL by AIC & 1.00 \\
Mean $\Delta {\rm BIC}$ & $-34.692$ \\
Median $\Delta {\rm BIC}$ & $-33.263$ \\
Fraction favouring CPL by BIC & 1.00 \\
\hline
\end{tabular}
\end{table}

We also compute ${\rm FoM}=1/\sqrt{\det\operatorname{Cov}(w_0,w_1)}$, which measures likelihood-region compactness rather than model preference. The present GRB sample gives 0.487. Three representative CPL-injected mocks near the 16th, 50th, and 84th percentiles of the $\Delta\chi^2$ distribution give 2.229, 3.841, and 2.528. Their mean FoM is 2.866, corresponding to a mean gain of 5.88, with mean one-dimensional errors $\langle\sigma(w_0)\rangle=0.309$ and $\langle\sigma(w_1)\rangle=0.839$. A representative $\Lambda$CDM mock has FoM 11.435; it is not the same ensemble as the three CPL-injected examples.

The forecasts illustrate the potential of a large homogeneous high-redshift distance sample. They remain intentionally idealized: luminosity functions, prompt-spectrum selection, triggering, follow-up efficiency, redshift completeness, correlated calibration systematics, and possible evolution are not simulated. The assigned standardized-distance precision and fiducial cosmology are conditions of the result, not demonstrated mission performance. A population- and instrument-level forecast is needed before interpreting these values as predictions for THESEUS itself.

\FloatBarrier
\section{Conclusions}\label{sec:conclusions}
We have constructed a Pantheon+-anchored GRB calibration using a Bayesian neural network with explicitly assessed and recalibrated predictive uncertainties. The low-redshift Amati fit yields a Hubble diagram extending to $z\simeq9.4$. GRBs thereby provide a complementary redshift lever arm, while their individual errors remain limited by intrinsic correlation scatter and sample heterogeneity.

The D'Agostini and Reichart prescriptions give different slopes, with the latter defining the supplied baseline diagram. Controlled full-covariance local-regression tests quantify SN--GRB dependence and sensitivity to the three highest-redshift supernovae. Their effects on posterior medians and widths are small, but these surrogate diagnostics do not replace BNN retraining or direct propagation of its training-level covariance.

The calibration sample does not require a redshift term orthogonal to the Amati relation. The effective-truncation EP test nevertheless finds positive marginal isotropic-energy evolution, compatible with evolution along the relation. This distinction reconciles the parametric and modified-rank diagnostics while preserving the limitation imposed by an approximate selection function.

Combining the calibrated GRBs with Pantheon+, observational Hubble data, and DESI BAO improves phenomenological fits with dark-energy extensions. AIC favours CPL while BIC favours constant-$w$ dark energy, so the present comparison is not conclusive evidence for an evolving equation of state. Under explicitly idealized assumptions, future standardized-GRB catalogues have substantial model-discrimination power. More homogeneous observations and a complete treatment of selection and calibration systematics are required to turn that potential into robust high-redshift cosmological tests.

\begin{acknowledgements}
EP acknowledges the support of INFN Sez. di Napoli (Iniziative Specifiche QGSKY and InDark).
\end{acknowledgements}
\section*{Data availability}
The calibrated GRB Hubble diagram is supplied in the machine-readable file \texttt{grb\_hubble\_diagram\_B2\_full.dat}, containing redshift, distance modulus, and uncertainty. Appendix~\ref{app:data} reproduces the complete supplied numerical table. The remaining GRB catalogue and numerical outputs are available from the corresponding author upon reasonable request. Pantheon+ and DESI BAO data are publicly available from their survey repositories.
\FloatBarrier
\label{main-text-end}
\bibliographystyle{aa}
\bibliography{biblio}

\clearpage
\label{appendix-start}
\appendix
\nolinenumbers
\section{Neural-network implementation and calibration diagnostics}\label{app:bnn}
\subsection{Variational settings and training}
For weights $\mathcal W$ and data $\mathcal D$, Bayesian prediction marginalizes the conditional output over $p(\mathcal W\mid\mathcal D)$. In the Concrete Dropout approximation \citep{ghk17}, stochastic masks provide samples from a variational weight distribution. We use $p$ consistently for a retention probability; the dropout probability is $1-p$. A relaxed mask for neuron $i$ is
\begin{equation}
\begin{split}
z_i&=\operatorname{sigmoid}\!\left[
\frac{\log p-\log(1-p)+\log u-\log(1-u)}{t}\right],\\
&\hspace{30mm}u\sim U(0,1).
\end{split}
\end{equation}
The temperature is $t=0.1$ and the prior length scale follows the inverse-training-sample-size prescription. The variational precision is $\tau=34$, estimated from the inverse residual variance of a reference $\Lambda$CDM fit. Exploratory tests over $5\leq\tau\leq34$ gave nearly unchanged reconstructions and calibration metrics within the investigated range; this precision choice does not fix the cosmological shape of the reconstructed distances.

The full-covariance negative log-likelihood is given in Eq.~\ref{eq:bnn-nll}. During the 5000-epoch warmup, we minimize
\begin{equation}
\mathcal L_{\rm warm}=\mathcal L_{\rm NLL}
+\eta\frac1N\sum_i w_i(y_i-\mu_i)^2,
\qquad w_i\propto\widehat p(z_i)^{-1},
\end{equation}
where $\widehat p$ is a kernel-density estimate of the redshift distribution and $\eta$ is annealed to zero. Subsequent optimization uses the likelihood alone. The log-determinant term penalizes inflated predictive variance as well as fitting the residuals. We use stochastic gradient descent with momentum. Training is halted when validation loss fails to improve beyond the mean of its preceding 30 evaluations for 3000 consecutive epochs. NLL and root-mean-square error are monitored throughout; Fig.~\ref{fig:bnn-diagnostics}a retains the supplied training curves.

The network outputs a heteroscedastic Gaussian predictive variance, which complements the released observational covariance. It is not identified with a particular physical noise source. Independently, $N_{\rm MC}=1000$ stochastic passes estimate epistemic variance. The two components are combined before uncertainty recalibration.

\subsection{Coverage, sharpness, and localized variance excursions}
\begin{figure*}
\centering
\begin{minipage}{0.48\textwidth}\centering
\includegraphics[width=\linewidth]{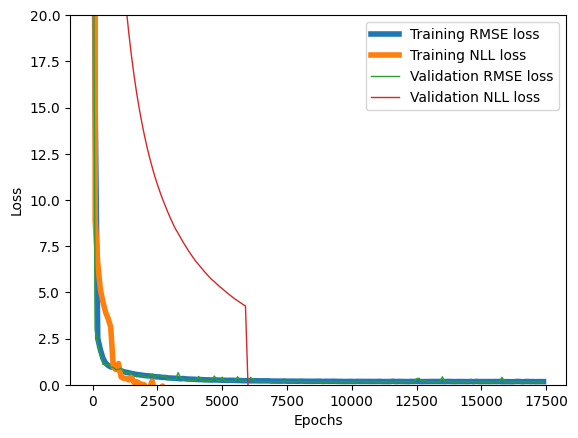}\\(a) Training and validation
\end{minipage}\hfill
\begin{minipage}{0.48\textwidth}\centering
\includegraphics[width=\linewidth]{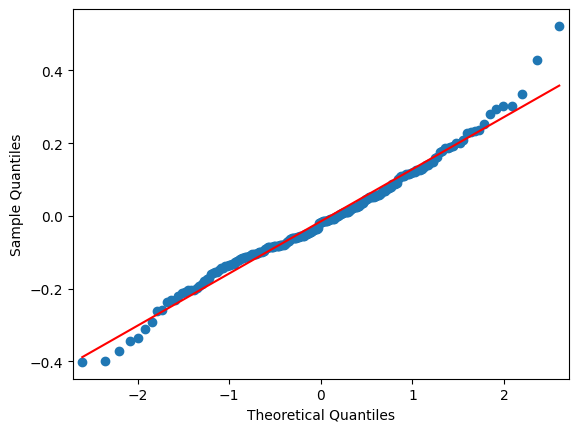}\\(b) Standardized-residual Q--Q diagnostic
\end{minipage}
\par\medskip
\begin{minipage}{0.48\textwidth}\centering
\includegraphics[width=\linewidth]{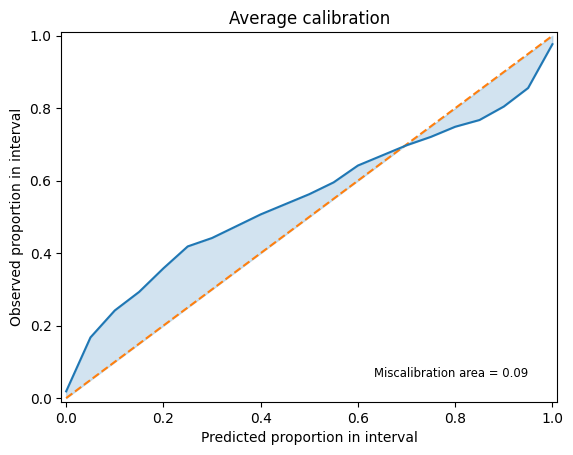}\\(c) Before uncertainty recalibration
\end{minipage}\hfill
\begin{minipage}{0.48\textwidth}\centering
\includegraphics[width=\linewidth]{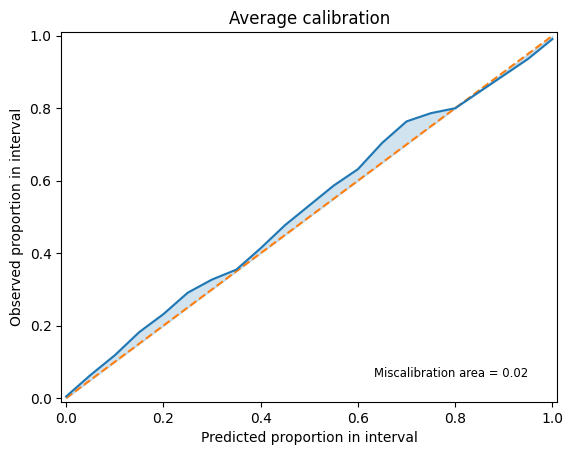}\\(d) After uncertainty recalibration
\end{minipage}
\caption{BNN diagnostics, retaining all four supplied plots. (a) Training and validation RMSE and Mahalanobis distance; the first 5000 epochs are the warmup. (b) Q--Q plot, whose red line fits the empirical quantiles. (c,d) Reliability diagrams: the dashed identity denotes ideal coverage, and the miscalibration area falls from 0.09 to 0.02.}
\label{fig:bnn-diagnostics}
\end{figure*}

For a predictive cumulative distribution $F_i$ and quantile $F_i^{-1}(p)$, empirical coverage is
\begin{equation}
\widehat F(p)=\frac1T\sum_{i=1}^T
\mathbf1\{y_i\leq F_i^{-1}(p)\}.
\end{equation}
A calibrated predictor has $\widehat F(p)\simeq p$. The diagnostics used here are
\begin{equation}
\mathcal A_{\rm mis}=\int_0^1|\widehat F(p)-p|\,dp,
\qquad
\mathcal S=\frac1T\sum_i\widehat\sigma_i^2.
\end{equation}
The miscalibration area is evaluated numerically over probability bins; the sharpness convention $\mathcal S$ is the mean variance. All reported metrics are averaged over five independently trained networks. The Uncertainty Toolbox routine \texttt{optimize\_recalibration\_ratio} fits a single multiplicative uncertainty scale on the dedicated calibration subset. It preserves relative uncertainty structure. The independent test diagnostics give $(\mathcal A_{\rm mis},\mathcal S)=(0.09,0.264)$ before and $(0.02,0.126)$ after recalibration. Figure~\ref{fig:bnn-diagnostics} shows the Gaussian Q--Q and reliability diagnostics; $z<0.02$ objects are excluded from their evaluation.

Repeated independent dropout evaluations occasionally produce localized epistemic-variance excursions whose positions vary between realizations. On a dense redshift grid, more than 87\% of the detected excursions are isolated points, with very few extending over more than three consecutive points. Before uncertainty propagation to the GRB calibration sample, estimates above 1.5 times their local median are replaced by that median. Only a small fraction of evaluated points is affected. Repetition with independent dropout draws changes the identities of affected GRBs, while the calibration and sharpness metrics remain unchanged within their statistical uncertainties. The adopted correction is therefore a numerical treatment of unstable variance excursions; it is part of the supplied uncertainty-processing procedure.

\section{Local-regression calibration and covariance diagnostics}\label{app:calibration}
\subsection{Local reconstruction and likelihoods}
In this subsection we compare the BNN-based calibration with an updated version of the local-regression technique, inspired by the standardization of SNIa with Cepheid variables and adopted in previous works \citep[see, for instance,][]{demianski21}, using the Pantheon+ sample. The structure of the local regression routine, can be schematized through the following points:
\begin{enumerate}
\item[$\star$]{set the redshift $z$ where $\mu(z)$ has to be reconstructed;}
\item[$\star$]{sort the mock SNIa sample  by increasing the value of $|z-z_i|$ and
select the first $n =\alpha {\cal{N}}_{SNIa}$, where $\alpha$ is a user-selected value and ${\cal{N}}_{SNIa}$ the total number of SNIa;}
\item[$\star$]{apply the weight function\,
\begin{equation}
W(u) = \left \{
\begin{array}{ll}
(1 - |u|^2)^2 & |u| \le 1 \\ ~ & ~ \\ 0 & |u| \ge 1
\end{array}\,,
\right.
\label{eq: wdef}
\end{equation}
}
where $u=|z - z_i|/\Delta$ and $\Delta$ the highest value of the $|z-z_i|$ over the previously selected subset;
\end{enumerate}
\begin{enumerate}
\item[$\star$]{fit a first-order polynomial to the data previously selected and weighted, and use the zeroth-order term as the best-fit value of the distance modulus  $\mu(z)$},
\item[$\star$]{evaluate the error $\sigma_{\mu}$ as the root mean square of the weighted residuals with respect to the best-fit value.}
\end{enumerate}
The neighbourhood size and polynomial order involve a compromise: the neighbourhood must contain enough SNe for a statistically meaningful fit, while remaining sufficiently local for a low-order polynomial to describe the distance–redshift relation.  
Once the regression provides an estimate of \(\mu(z)\), the reconstructed distance modulus can be used to calibrate the \(E_{p,i}\)–\(E_{\rm iso}\) relation. We considered only GRBs with $z \le 1.5$, consistently with the low-redshift calibration cut adopted in the baseline analysis.
To standardize the $E_{\rm p,i}$ -- $E_{\rm iso}$ relation as expressed
by  Eq. (\ref{correlation}), we need to
fit the data arrays $\{x_i, y_i\}$ with uncertainties $\{\sigma_{x_i}, \sigma_{y_i}\}$, to a straight line
\begin{equation}
\label{linear} y =b + a x\,,
\end{equation}
where the  parameters $(a, b)$ are fit parameters. Moreover, since the $E_{\rm p,i}$ -- $E_{\rm iso}$ correlation exhibits a certain amount of intrinsic extra scatter, $\sigma_{int}$. We expect that $\sigma_{int}$  is determined, together with 
\((a,b)\), by the fitting procedure itself. In this respect, we adopt a new approach,  defining a new full-covariance D'Agostini-type likelihood to include the full covariance matrix of the reconstructed
$\log_{10}(E_{\rm iso}/{\rm erg})$ values. The intrinsic scatter was
modelled in the dependent variable. For the
residual vector
\[
\mathbf{r}=\mathbf{y}-a\mathbf{x}-b\mathbf{1},
\]
the total covariance matrix was defined as
\[
\mathbf{V}(a,\sigma_{\rm int})
=
\mathbf{C}_{y}
+
\operatorname{diag}
\left(
a^{2}\sigma_{x,i}^{2}
+
\sigma_{\rm int}^{2}
\right),
\]
and the negative calibration log-likelihood was written as
\[
-2\ln\mathcal{L}
=
N\ln(2\pi)
+
\ln\det\mathbf{V}
+
\mathbf{r}^{T}\mathbf{V}^{-1}\mathbf{r}.
\]
No Reichart geometrical normalization factor proportional to
$(1+a^{2})^{N/2}$ was included at this stage.
The matrix $\mathbf{C}_y$ denotes the observational covariance matrix
of the reconstructed quantities
\begin{equation}
y_i =
\log_{10}\left(
\frac{E_{{\rm iso},i}}{\rm erg}
\right).
\end{equation}
Since
\begin{equation}
\log_{10} E_{{\rm iso},i}
=
{\rm const.}
+
\frac{2}{5}\mu_i
+
\log_{10} S_{{\rm bolo},i}
-
\log_{10}(1+z_i),
\end{equation}
the covariance between two reconstructed values $y_i$ and $y_j$ is
given by
\begin{equation}
\left(\mathbf{C}_y\right)_{ij}
=
\left(\frac{2}{5}\right)^2
\left(\mathbf{C}_{\mu}\right)_{ij}
+
\delta_{ij}
\left[
\frac{\sigma_{S_{{\rm bolo},i}}}
     {S_{{\rm bolo},i}\ln 10}
\right]^2,
\end{equation}
where $\delta_{ij}$ is the Kronecker delta and
$\mathbf{C}_{\mu}$ is the covariance matrix of the distance moduli
reconstructed from the Pantheon+ sample.
The latter is obtained by propagating the full Pantheon+ covariance
matrix through the local-regression estimator:
\begin{equation}
\mathbf{C}_{\mu}
=
\mathbf{A}\,
\mathbf{C}_{\rm SN}\,
\mathbf{A}^{T},
\end{equation}
where $\mathbf{C}_{\rm SN}$ is the full statistical-plus-systematic
Pantheon+ covariance matrix and $\mathbf{A}$ is the matrix of linear
coefficients defining the local distance-modulus reconstruction,
\begin{equation}
\widehat{\mu}_i
=
\sum_{\alpha}
A_{i\alpha}\,
\mu_{\alpha}^{\rm SN}.
\end{equation}
Therefore, in compact matrix notation,
\begin{equation}
\mathbf{C}_y
=
\left(\frac{2}{5}\right)^2
\mathbf{A}\,
\mathbf{C}_{\rm SN}\,
\mathbf{A}^{T}
+
\operatorname{Diag}
\left[
\left(
\frac{\sigma_{S_{{\rm bolo},i}}}
     {S_{{\rm bolo},i}\ln 10}
\right)^2
\right].
\end{equation}
For comparison with the other calibration approaches, we also consider a full-covariance Reichart-type likelihood, which could provide useful context for the inferred Amati-relation parameters, we also include a Reichart geometrical normalization factor proportional to
$(1+a^{2})^{N/2}$, but accounting for the
non-diagonal covariance of the reconstructed isotropic energies. The negative log-likelihood is then
\begin{equation}
-2\ln\mathcal{L}_{\mathrm{R}}
=
N\ln(2\pi)
+
\ln\det\mathbf{V}
+
\mathbf{r}^{T}\mathbf{V}^{-1}\mathbf{r}
-
N\ln(1+a^{2}).
\end{equation}
The first three terms define the usual multivariate Gaussian errors-in-variables contribution, whereas the last term is the Reichart geometrical normalization, which depends explicitly on the slope and distinguishes the Reichart prescription from the
corresponding D'Agostini-type likelihood. It is important to stress that this expression should be regarded as a pragmatic full-covariance Reichart-type prescription, rather than as a unique extension of the original Reichart likelihood. The latter was derived
for independent data points, whereas in the present local-regression calibration the reconstructed values of \(\log E_{\rm iso}\) are correlated because they inherit the Pantheon+ covariance through the reconstruction operator. We therefore replace the independent Gaussian residual term by the
corresponding multivariate Gaussian form, while retaining the slope-dependent geometrical normalization factor that characterizes the Reichart approach. This choice allows us to test the impact of a
Reichart-like slope normalization in the presence of correlated reconstructed distance moduli.
Moreover, it is worth noting that the full-covariance Reichart-type likelihood is used here only as a
diagnostic and comparison tool. In particular, it does not define the baseline GRB Hubble diagram used for the cosmological constraints, and the main cosmological conclusions do not rely on the uniqueness of this generalization.
We explore the posterior using three parallel chains with flat priors on \(a\), \(b\), and \(\sigma_{\rm int}\). Convergence is monitored using the Gelman–Rubin statistic, requiring \(R-1\simeq0.01\) for every parameter. We discard the first 30\% of each chain as burn-in, thin the remaining samples, and merge the thinned chains. Posterior medians and credible intervals are then calculated from the combined samples.

\begin{table}
\centering
\caption{
Posterior constraints on the calibration parameters of the
\(E_{\rm p,i}\)--\(E_{\rm iso}\) correlation obtained from the local
regression (LR) and Bayesian neural-network (BNN) reconstructions.
For each calibration, the adopted likelihood prescription is reported.
}
\label{tab:ep-eiso-calibration}
\renewcommand{\arraystretch}{1.05}
\setlength{\tabcolsep}{ 2 pt}
\small
\resizebox{\columnwidth}{!}{%
\begin{tabular}{llccc}
\hline
\hline
Reconstruction
& Likelihood
& \(b\)
& \(a\)
& \(\sigma_{\rm int}\)
\\
\hline
LR--Pantheon+
& Full-cov
& \(52.80^{+0.04}_{-0.04}\)
& \(1.66^{+0.09}_{-0.09}\)
& \(0.43^{+0.03}_{-0.03}\)
\\
LR--Pantheon+
& Reichart full-cov
& \(52.84^{+0.05}_{-0.05}\)
& \(2.13^{+0.11}_{-0.11}\)
& \(0.47^{+0.04}_{-0.04}\)
\\
BNN--Pantheon+
& Reichart
& \(52.78^{+0.09}_{-0.09}\)
& \(2.13^{+0.24}_{-0.20}\)
& \(0.472^{+0.097}_{-0.072}\)
\\
BNN--Pantheon+
& D'Agostini & $52.83_{-0.08}^{+0.08}$  & $1.67_{-0.17}^{+0.17}$ & $0.436^{+0.085}_{-0.075}$ \\
LR--mock sample
& Full-cov
& \(52.60 \pm 0.06\)
& \(1.67 \pm 0.08\)
& \(0.43 \pm 0.03\)
\\
\hline
\hline
\end{tabular}%
}
\vspace{0.15cm}
\begin{minipage}{0.92\linewidth}
\footnotesize
Notes. ``Full-cov'' denotes the D'Agostini-type calibration including the
full covariance matrix of the reconstructed \(\log E_{\rm iso}\) values.
``Reichart full-cov'' denotes the corresponding full-covariance
generalization including the Reichart slope-dependent geometrical
normalization factor.
\end{minipage}
\end{table}
\begin{table}
    \caption{Calibration results of $E_{\rm p,i} $ - $E_{\rm iso} $ correlation  as reported in \citet{xie25}, using the J221 sample and a redshift cut-off of $z = 1.4$.  }
    \label{tab:amati_params_j221}
    \renewcommand{\arraystretch}{1.15}
    \centering
    \begin{tabular}{lccc}
        \hline
        \hline
        Method & \( b \) & \( a \) & \( \sigma_{int} \) \\
        \hline
        \noalign{\vskip 2mm}
        D'Agostini & $52.83_{-0.58}^{+0.58}$  & $1.59_{-0.10}^{+0.10}$ & $0.433$ \\
        Reichart   & $52.87_{-0.58}^{+0.58}$ & $2.026_{-0.093}^{+0.083}$ & $0.423$ \\
        \noalign{\vskip 2mm}
        \hline
        \hline
    \end{tabular}
\end{table}

\subsection{Cross-covariance and leave-three-out diagnostics}\label{app:sensitivity}
Let $\mathbf C_{\rm SN,GRB}$ be the SN--GRB cross-covariance obtained by propagating the implicit likelihood response of the Reichart calibration through the local-regression operator. The shared and conditional GRB covariances are
\begin{align}
\mathbf C_{\rm shared}&=\mathbf C_{\rm SN,GRB}^T
\mathbf C_{\rm SN}^{-1}\mathbf C_{\rm SN,GRB},\\
\mathbf S_{\rm GRB}&=\mathbf C_{\rm GRB}-\mathbf C_{\rm shared}.
\end{align}
The shared component is positive semidefinite within numerical precision. The Schur complement has $\lambda_{\min}=1.44875$, and $\lambda_{\max}(\mathbf C_{\rm shared})/\lambda_{\max}(\mathbf C_{\rm GRB})=0.0534$. Comparing block and joint likelihoods over 2800 converged posterior samples gives $\operatorname{median}(\Delta\chi^2)=-1.252$ with standard deviation 0.217. The normalized median displacement is defined by
\begin{equation}
\Delta_{\rm norm}(\theta)=
\frac{\theta_{\rm joint}^{\rm med}-\theta_{\rm block}^{\rm med}}
{\sqrt{\sigma_{\rm joint}^2+\sigma_{\rm block}^2}},
\qquad\sigma=\frac{q_{84}-q_{16}}2.
\end{equation}
The largest magnitude is $0.040\sigma$; values for $\Omega_m$, $w_a$, and $h$ are approximately 0.031, 0.0023, and 0.0011 in these units. The largest relative 68\% width change is 2.71\% for $w_0$.

For the leave-three-out test, $\alpha=0.02$ corresponds to approximately 34 neighbouring supernovae. We fix $n=34$ in the primary comparison to separate the effect of removing objects from that of changing neighbourhood membership through rounding; the nominal-fraction reconstruction is an additional algorithmic check. Of the 123 calibration targets, 20 change, all between redshifts 1.293 and 1.489. The maximum absolute shift is 0.02338 mag, approximately $1\sigma$. The largest standardized shift, $1.76\sigma$, instead corresponds to an absolute displacement of only 0.00193 mag.

Refitting gives $\Delta a=1.04\times10^{-4}$, $\Delta b=4.16\times10^{-4}$, and $\Delta\sigma_{\rm int}=8.84\times10^{-5}$. The full diagram changes by at most $1.31\times10^{-3}$ mag, or $1.0\times10^{-3}\sigma$. Its relative covariance Frobenius-norm change is $3.19\times10^{-4}$ and the largest individual uncertainty change is 0.0183\%; the covariance remains positive definite. Importance reweighting has ${\rm ESS}/N\simeq0.88$, no resolved median shift, and a largest 68\% width change of 0.129\%. These tests are controlled local-regression diagnostics because exact retraining products for the baseline BNN are unavailable.

\section{Efron--Petrosian analysis details}\label{app:evolution}
\label{app:ep-details}
\subsection{Associated sets and modified Kendall statistic}
For each GRB we define
\begin{equation}
y_i=\log_{10}\!\left(\frac{E_{{\rm iso},i}}{{\rm erg}}\right),
\qquad
u_i=\log_{10}(1+z_i),
\end{equation}
and, for a trial evolution index $k$,
\begin{equation}
y'_i(k)=y_i-ku_i,
\qquad
y'_{\lim,i}(k)=y_{\lim,i}-ku_i,
\end{equation}
where
\begin{equation}
y_{\lim,i}=\log_{10}\!\left[
\frac{4\pi d_L^2(z_i)S_{{\rm bolo},\lim}}{(1+z_i){\rm erg}}
\right].
\end{equation}
For object $i$, the associated set is
\begin{equation}
\label{eq:ep-associated-set}
{\cal A}_i(k)=
\left\{j:\ y'_j(k)\geq y'_i(k),\quad
             y'_{\lim,j}(k)\leq y'_i(k)\right\}.
\end{equation}
The first condition selects objects at least as energetic as object $i$ after de-evolution, while the second requires that an object with de-evolved energy $y'_i$ would remain observable at the redshift of object $j$. The object itself belongs to its associated set. Let
\begin{equation}
N_i=|{\cal A}_i|
\end{equation}
be the associated-set size and $R_i$ the average rank of $z_i$ among the redshifts in ${\cal A}_i$. Under the null hypothesis that $y'(k)$ and redshift are independent within the truncated sample, the expectation and variance of the rank are
\begin{equation}
\label{eq:ep-EV}
E_i=\frac{N_i+1}{2},
\qquad
V_i=\frac{N_i^2-1}{12}.
\end{equation}
The Efron--Petrosian modified Kendall statistic is
\begin{equation}
\label{eq:ep-tau-full}
\tau_{\rm EP}(k)=
\frac{\displaystyle\sum_{i:N_i>1}(R_i-E_i)}
     {\displaystyle\left[\sum_{i:N_i>1}V_i\right]^{1/2}}.
\end{equation}
The best-fit evolution index satisfies $\tau_{\rm EP}(k_{\rm iso})=0$, and the crossings with $\tau_{\rm EP}=\pm1$ define the conventional 68\% interval. For the 10\% baseline sample, 237 of the 238 objects contribute, the associated-set sizes range from 2 to 157, and their median is 94.
\subsection{Detailed effective-threshold scans}
The baseline effective fluence threshold was chosen at the 10th percentile of the observed distribution. To assess sensitivity to this choice, the complete calculation was repeated at the 5th and 7th percentiles. The selected samples and inferred parameters are reported in Table~\ref{tab:ep-threshold-full}.
\begin{table*}
\centering
\caption{Detailed Efron--Petrosian results for the effective bolometric-fluence threshold scan. The quoted lower and upper limits are the crossings of $\tau_{\rm EP}=\pm1$.}
\label{tab:ep-threshold-full}
\small
\begin{tabular}{cccccccc}
\toprule
Percentile & $S_{{\rm bolo},\lim}$ [${\rm erg\,cm^{-2}}$] & Retained & Removed & $\tau_{\rm EP}(0)$ & $k_{\rm iso}$ & Lower 68\% & Upper 68\% \\
\midrule
5\%  & $1.20524\times10^{-6}$ & 251 & 13 (4.92\%) & $-3.58830$ & $1.83095^{+0.48247}_{-0.54301}$ & 1.28794 & 2.31342 \\
7\%  & $1.50854\times10^{-6}$ & 246 & 18 (6.82\%) & $-3.67996$ & $1.98146^{+0.40969}_{-0.59442}$ & 1.38703 & 2.39115 \\
10\% & $2.26269\times10^{-6}$ & 238 & 26 (9.85\%) & $-3.18660$ & $1.71401^{+0.71732}_{-0.55540}$ & 1.15860 & 2.43132 \\
\bottomrule
\end{tabular}
\end{table*}
The three intervals overlap strongly. Their central values differ by less than the corresponding statistical errors, so the indication of positive marginal evolution is not driven by a single percentile choice.
\subsection{Dependence on the fiducial matter density}
The observed fluences and the effective threshold were held fixed while $E_{\rm iso}$ and $E_{{\rm iso},\lim}$ were recomputed in flat $\Lambda$CDM for $0.20\leq\Omega_m\leq0.40$. Table~\ref{tab:ep-cosmology-full} gives the complete scan. The variation of the central value over the full interval is $\Delta k_{\rm iso}=0.215$, smaller than the nominal EP uncertainty. A change in $H_0$ produces the same additive shift in $\log E_{\rm iso}$ and $\log E_{{\rm iso},\lim}$ and therefore leaves the associated sets and $\tau_{\rm EP}$ unchanged.
\begin{table*}
\centering
\caption{Complete stability scan with respect to the fiducial flat-$\Lambda$CDM matter density.}
\label{tab:ep-cosmology-full}
\small
\begin{tabular}{ccccccccc}
\toprule
$\Omega_m$ & $\Omega_\Lambda$ & $\tau_{\rm EP}(0)$ & $k_{\rm iso}$ & $-\sigma$ & $+\sigma$ & Lower 68\% & Upper 68\% & $\tau_{\rm EP}(k_{\rm iso})$ \\
\midrule
0.20 & 0.80 & $-3.38046$ & 1.84767 & 0.54415 & 0.75511 & 1.30352 & 2.60277 & $-1.85\times10^{-3}$ \\
0.25 & 0.75 & $-3.24657$ & 1.77350 & 0.55728 & 0.74315 & 1.21622 & 2.51664 & $ 7.38\times10^{-3}$ \\
0.30 & 0.70 & $-3.18660$ & 1.71401 & 0.55540 & 0.71732 & 1.15860 & 2.43132 & $-4.61\times10^{-3}$ \\
0.35 & 0.65 & $-3.09147$ & 1.67704 & 0.55610 & 0.69774 & 1.12095 & 2.37478 & $ 9.20\times10^{-4}$ \\
0.40 & 0.60 & $-3.03393$ & 1.63293 & 0.55282 & 0.69056 & 1.08011 & 2.32349 & $ 9.19\times10^{-4}$ \\
\bottomrule
\end{tabular}
\end{table*}
\subsection{Monte Carlo treatments of fluence uncertainties and threshold membership}
Fluence uncertainties were represented by positive lognormal distributions whose means and standard deviations reproduce the catalogue values. Two conceptually different experiments were performed.
In the \emph{variable-membership} experiment, the effective threshold was reapplied after each fluence realization. This propagates both measurement errors and changes in sample composition. All 1000 realizations were successful and yielded
\begin{equation}
k_{\rm iso}^{\rm var}=1.76908^{+0.17012}_{-0.16155},
\qquad
\sigma_{\rm var}=0.16975.
\end{equation}
The selected sample ranged from 231 to 242 GRBs, with median 237. The recovered evolution parameter is strongly anticorrelated with the number of selected objects,
\begin{equation}
\rho_{\rm P}(N_{\rm sel},k_{\rm iso})=-0.647,
\qquad
\rho_{\rm S}(N_{\rm sel},k_{\rm iso})=-0.637.
\end{equation}
For the 201 realizations containing exactly 238 objects, the median is $k_{\rm iso}=1.71493$, essentially identical to the nominal result. The variable-membership experiment is therefore interpreted as a threshold-membership sensitivity test, not as a pure measurement-error propagation.
In the \emph{fixed-sample, threshold-conditioned} experiment, the nominal 238 GRBs were retained in every realization and their lognormal fluence distributions were conditioned to remain above $S_{{\rm bolo},\lim}$. This isolates the sensitivity to observational errors conditional on the adopted sample. The final 1000-realization summary is
\begin{equation}
\label{eq:ep-cond-mc}
k_{\rm iso}^{\rm cond}=1.73516^{+0.11484}_{-0.09575},
\qquad
\sigma_{\rm cond}=0.10770,
\end{equation}
with central 95\% interval $1.56758<k_{\rm iso}^{\rm cond}<1.99347$. The median shift relative to the nominal EP solution is only 0.02208. The conditional spread is therefore much smaller than the asymmetric EP interval and is quoted separately rather than combined in quadrature.
\begin{table*}
\centering
\caption{Summary of the two 1000-realization Monte Carlo experiments.}
\label{tab:ep-mc-full}
\small
\begin{tabular}{lcccccccc}
\toprule
Experiment & Successes & Mean & Median & Std. dev. & 16th & 84th & 2.5th & 97.5th \\
\midrule
Variable membership & 1000 & 1.77322 & 1.76908 & 0.16975 & 1.60753 & 1.93920 & 1.46641 & 2.14375 \\
Fixed and conditioned & 1000 & 1.74651 & 1.73516 & 0.10770 & 1.63940 & 1.85000 & 1.56758 & 1.99347 \\
\bottomrule
\end{tabular}
\end{table*}
\subsection{Bootstrap precision and numerical zero diagnostics}
A 5000-replicate non-parametric bootstrap of the 1000 variable-membership values was used to assess the numerical precision of the Monte Carlo summaries. The standard error of the direct Monte Carlo mean is 0.00537, consistent with the bootstrap error 0.00526. The bootstrap errors on the median, 16th percentile, and 84th percentile are 0.00531, 0.00729, and 0.01256, respectively. Hence 1000 realizations are sufficient for the reported precision.

The numerical root finding was also explicitly checked. For the nominal $\Omega_m=0.3$ solution, $\tau_{\rm EP}(k_{\rm iso})=-4.61\times10^{-3}$. Across the variable-membership and conditional Monte Carlo samples, the maximum absolute residual statistics at the reconstructed zeros were $2.77\times10^{-3}$ and $2.81\times10^{-3}$, respectively. In the variable-membership calculation the maximum final root bracket width was $2.44\times10^{-5}$; the median width was zero because many realizations met the stopping criterion exactly at the numerical precision adopted. These values confirm that numerical root-finding errors are negligible compared with the statistical uncertainties.

\section{Future-sample likelihood and complete forecast results}\label{app:forecast}
The future mock observables and assigned uncertainties are defined in Eq.~\ref{eq:mock-observable}. To remove the unknown absolute distance zero point, the likelihood includes a free offset:
\begin{equation}
\chi^2(\theta,\Delta\mu)=\sum_i
\frac{[\mu_i-\mu_{\rm th}(z_i;\theta)-\Delta\mu]^2}{\sigma_{\mu,i}^2}.
\end{equation}
For $r_i=\mu_i-\mu_{\rm th}(z_i;\theta)$, define
\begin{equation}
A=\sum_i\sigma_{\mu,i}^{-2},\qquad
B=\sum_i r_i\sigma_{\mu,i}^{-2},\qquad
C=\sum_i r_i^2\sigma_{\mu,i}^{-2}.
\end{equation}
Then $\widehat{\Delta\mu}=B/A$ and $\chi^2_{\rm prof}=C-B^2/A$. The fitted parameter counts are $K=2$ for $\Lambda$CDM, $(\Omega_m,\Delta\mu)$, and $K=4$ for CPL, $(\Omega_m,w_0,w_1,\Delta\mu)$. The profiled offset is included when computing AIC and BIC. We use $\Delta\chi^2=\chi^2_{\Lambda\rm CDM}-\chi^2_{\rm CPL}$, and the two-parameter thresholds 2.30, 6.17, and 11.8 for nominal $1\sigma$, $2\sigma$, and $3\sigma$ comparisons. The forecast information-criterion differences are CPL minus $\Lambda$CDM, so negative values favour CPL; this sign convention differs from the minimum-relative convention in the observed-data comparison.

Tables~\ref{tab:lcdm_null_test}, \ref{tab:cpl_detection_forecast}, \ref{tab:cpl_recovery}, and \ref{tab:cpl_fom_three_mocks} retain the supplied null-test, detection, parameter-recovery, and likelihood-FoM results. FoM ratios use the present-GRB value 0.487. Across the three representative CPL-injected likelihood grids, mean, median, minimum, and maximum FoM are 2.866, 2.528, 2.229, and 3.841; mean and median gains are 5.88 and 5.19. These examples characterize the adopted mocks and do not include an instrument-level or population-level selection calculation.

\begin{table}
\centering
\caption{Parameter recovery and boundary diagnostics for the 50 CPL-injected mock catalogues. Biases are defined as recovered minus fiducial.}
\label{tab:cpl_recovery}
\begin{tabular}{lc}
\hline
Quantity & Value \\
\hline
Mean bias in $\Omega_m$ & 0.0076 \\
Mean bias in $w_0$ & 0.238 \\
Mean bias in $w_1$ & $-1.257$ \\
Std. dev. of recovered $\Omega_m$ & 0.0286 \\
Std. dev. of recovered $w_0$ & 0.460 \\
Std. dev. of recovered $w_1$ & 3.006 \\
Fraction at lower $\Omega_m$ bound & 0.00 \\
Fraction at upper $\Omega_m$ bound & 0.00 \\
Fraction at lower $w_0$ bound & 0.00 \\
Fraction at upper $w_0$ bound & 0.00 \\
Fraction at lower $w_1$ bound & 0.06 \\
Fraction at upper $w_1$ bound & 0.00 \\
\hline
\end{tabular}
\end{table}
\begin{table*}
\centering
\caption{Likelihood-based FoM for three representative CPL-injected future mock
catalogues.  The three mocks are selected near the 16th, 50th, and 84th
percentiles of the $\Delta\chi^2$ distribution.}
\label{tab:cpl_fom_three_mocks}
\begin{tabular}{lcccc}
\hline
Mock & Seed & $\Delta\chi^2$ & FoM  & FoM/FoM$_{\rm real}$  \\
\hline
Low-$\Delta\chi^2$ & 96022 & 36.969 & 2.229  & 4.58  \\
Median-$\Delta\chi^2$ & 96041 & 47.716 & 3.841   & 7.88  \\
High-$\Delta\chi^2$ & 96019 & 59.645 & 2.528  & 5.19  \\
\hline
\end{tabular}
\end{table*}

\section{Information criteria and tabulated data}\label{app:details}
Equation~\ref{aic} gives the large-sample AIC and BIC. The finite-sample correction is ${\rm AIC}_{\rm c}={\rm AIC}+2K(K+1)/(N-K-1)$. For observed-data model $i$, $\Delta{\rm IC}_i={\rm IC}_i-\min_j{\rm IC}_j$. As qualitative guides, $\Delta{\rm AIC}\lesssim2$ indicates substantial relative support, whereas $\Delta{\rm AIC}\gtrsim10$ indicates little support. BIC differences below 2 are weak, 2--6 positive, 6--10 strong, and above 10 very strong evidence against a model. These are descriptive thresholds, not sharp probability boundaries.

\subsection{Hubble measurements and GRB table}\label{app:data}
Table~\ref{Hz-data} reproduces the 34 observational Hubble measurements compiled by \citet{alm24}. The cosmological likelihood uses the correlated block described in Sect.~\ref{sec:cosmology}, rather than treating every entry as independent. Tables~\ref{calibratedHd_full} and \ref{calibratedHd_full_cont} reproduce all 264 rows of the calibrated GRB diagram. The complete numerical file \texttt{grb\_hubble\_diagram\_B2\_full.dat} accompanies both source packages.

\begin{table}
\caption{The OHD sample used in our analysis, compiled in \citet{alm24}. The columns are the redshifts, the $H(z)$ measurements, the statistical and  systematic uncertainties.}
\label{Hz-data}
\centering
\begin{tabular}{cccc}

\hline

$z$ & $H(z)$ & $\sigma_{H}^{stat}$ & $\sigma_{H}^{sys}$ \\
\hline \hline

 0.07 & 69. & 19.6 & 12.4 \\
 0.09 & 69. & 12. & 11.4 \\
 0.12 & 68.6 & 26.2 & 11.4 \\
 0.17 & 83. & 8. & 13.1 \\
 0.1791 & 75. & 3.8 & 0.5 \\
 0.1993 & 75. & 4.9 & 0.6 \\
 0.2 & 72.9 & 29.6 & 11.5 \\
 0.27 & 77. & 14. & 12.1 \\
 0.28 & 88.8 & 36.6 & 13.2 \\
 0.3519 & 83. & 13. & 4.8 \\
 0.3802 & 83. & 4.3 & 12.9 \\
 0.4 & 95. & 17. & 12.7 \\
 0.4004 & 77. & 2.1 & 10. \\
 0.4247 & 87.1 & 2.4 & 11. \\
 0.4497 & 92.8 & 4.5 & 12.1 \\
 0.47 & 89. & 23. & 44. \\
 0.4783 & 80.9 & 2.1 & 8.8 \\
 0.48 & 97. & 62. & 12.7 \\
 0.5929 & 104. & 11.6 & 4.5 \\
 0.6797 & 92. & 11.6 & 4.5 \\
 0.75 & 98.8 & 33.6 & 0 \\
 0.7812 & 105. & 9.4 & 6.1 \\
 0.8 & 113.1 & 15.1 & 20.2 \\
 0.8754 & 125. & 15.3 & 6. \\
 0.88 & 90. & 40. & 10.1 \\
 0.9 & 117. & 23. & 13.1 \\
 1.037 & 154. & 13.6 & 14.9 \\
 1.26 & 135. & 65. & 0 \\
 1.3 & 168. & 17. & 14. \\
 1.363 & 160. & 33.6 & 0 \\
 1.43 & 177. & 18. & 14.8 \\
 1.53 & 140. & 14. & 11.7 \\
 1.75 & 202. & 40. & 16.9 \\
 1.965 & 186.5 & 50.4 & 0 \\

 \hline
 \end{tabular}

\end{table}
\begin{table*}[p]
\centering
\caption{Complete supplied GRB Hubble diagram (rows 1--132). Each group of three columns lists redshift, distance modulus, and its uncertainty. Rows follow the supplied file; no numerical values have been changed.}
\label{calibratedHd_full}
\small
\begin{tabular}{rrr@{\hspace{14mm}}rrr@{\hspace{14mm}}rrr}
\toprule
$z$ & $\mu$ & $\sigma_\mu$ & $z$ & $\mu$ & $\sigma_\mu$ & $z$ & $\mu$ & $\sigma_\mu$ \\
\midrule
0.03351 & 34.442 & 1.914 & 0.73600 & 45.259 & 1.310 & 1.15500 & 44.893 & 1.511 \\
0.07580 & 36.858 & 2.068 & 0.73600 & 42.667 & 1.229 & 1.15880 & 46.302 & 1.462 \\
0.11700 & 40.151 & 1.244 & 0.74810 & 43.983 & 1.435 & 1.16000 & 43.964 & 1.279 \\
0.12540 & 39.571 & 1.310 & 0.78000 & 41.029 & 1.361 & 1.17000 & 43.255 & 1.217 \\
0.16850 & 38.789 & 1.350 & 0.80000 & 43.563 & 1.238 & 1.17150 & 44.533 & 1.187 \\
0.25000 & 39.131 & 2.350 & 0.80400 & 43.628 & 1.200 & 1.20760 & 43.715 & 1.192 \\
0.29300 & 40.052 & 2.070 & 0.80490 & 45.503 & 1.219 & 1.21000 & 45.255 & 1.188 \\
0.31000 & 40.288 & 1.325 & 0.80700 & 43.686 & 1.223 & 1.22000 & 43.204 & 1.184 \\
0.31300 & 43.891 & 1.207 & 0.82000 & 43.767 & 1.187 & 1.23800 & 42.809 & 1.250 \\
0.32850 & 39.484 & 1.191 & 0.83500 & 44.672 & 1.382 & 1.24000 & 43.673 & 1.256 \\
0.33990 & 42.105 & 1.317 & 0.83970 & 45.120 & 1.355 & 1.24600 & 44.819 & 1.210 \\
0.35000 & 38.013 & 1.554 & 0.84200 & 43.601 & 1.363 & 1.25250 & 44.548 & 1.310 \\
0.36000 & 44.151 & 1.432 & 0.84630 & 44.517 & 1.226 & 1.25400 & 43.635 & 1.377 \\
0.36700 & 42.255 & 1.235 & 0.85700 & 43.034 & 1.184 & 1.25600 & 45.005 & 1.260 \\
0.39000 & 40.282 & 1.278 & 0.85900 & 43.483 & 1.450 & 1.26220 & 44.296 & 1.254 \\
0.41000 & 41.053 & 1.468 & 0.87600 & 43.185 & 1.228 & 1.29300 & 44.127 & 1.196 \\
0.41400 & 43.717 & 1.501 & 0.88600 & 45.114 & 1.219 & 1.29700 & 42.883 & 1.225 \\
0.42450 & 43.040 & 1.257 & 0.89690 & 45.312 & 1.228 & 1.30000 & 43.851 & 1.310 \\
0.43400 & 42.911 & 1.373 & 0.93820 & 43.887 & 1.300 & 1.31450 & 46.799 & 1.316 \\
0.45000 & 40.760 & 1.191 & 0.94700 & 42.479 & 1.366 & 1.32900 & 45.291 & 1.233 \\
0.45000 & 42.078 & 1.361 & 0.95800 & 43.972 & 1.275 & 1.33100 & 46.877 & 1.698 \\
0.47910 & 41.678 & 1.312 & 0.96600 & 45.183 & 1.215 & 1.36800 & 46.050 & 1.308 \\
0.49000 & 40.389 & 1.323 & 0.97100 & 42.917 & 1.661 & 1.39800 & 43.870 & 1.241 \\
0.52950 & 42.899 & 1.413 & 0.97200 & 44.655 & 1.390 & 1.40500 & 46.258 & 1.424 \\
0.54000 & 41.089 & 1.187 & 0.98400 & 41.986 & 1.225 & 1.40600 & 43.874 & 1.312 \\
0.54200 & 41.427 & 1.213 & 1.01000 & 44.228 & 1.517 & 1.42300 & 45.235 & 1.343 \\
0.54300 & 42.537 & 1.693 & 1.02000 & 43.314 & 1.301 & 1.44000 & 45.183 & 1.187 \\
0.54400 & 42.472 & 1.191 & 1.02700 & 43.899 & 1.196 & 1.44360 & 44.233 & 1.310 \\
0.55000 & 42.904 & 1.248 & 1.03010 & 44.276 & 1.270 & 1.44500 & 44.601 & 1.187 \\
0.55700 & 44.333 & 1.191 & 1.03600 & 45.420 & 1.508 & 1.45200 & 45.559 & 1.354 \\
0.59700 & 43.569 & 1.250 & 1.05850 & 46.439 & 1.397 & 1.45200 & 46.795 & 1.243 \\
0.60600 & 41.967 & 1.225 & 1.06300 & 44.656 & 1.242 & 1.46000 & 44.293 & 1.263 \\
0.61800 & 44.355 & 1.342 & 1.09200 & 43.995 & 1.192 & 1.47680 & 44.205 & 1.228 \\
0.64390 & 41.984 & 1.182 & 1.09600 & 42.642 & 1.261 & 1.48000 & 43.000 & 1.728 \\
0.65350 & 41.720 & 1.580 & 1.09600 & 46.101 & 1.568 & 1.48900 & 45.990 & 1.328 \\
0.65400 & 42.698 & 1.235 & 1.10000 & 44.663 & 1.362 & 1.50260 & 46.581 & 1.343 \\
0.68870 & 44.781 & 1.196 & 1.10000 & 43.928 & 1.227 & 1.50500 & 41.718 & 1.422 \\
0.68980 & 42.790 & 1.208 & 1.10500 & 45.714 & 1.272 & 1.51690 & 46.818 & 1.267 \\
0.69500 & 43.801 & 1.387 & 1.11000 & 44.032 & 1.634 & 1.52000 & 43.395 & 1.223 \\
0.70550 & 42.062 & 1.187 & 1.11750 & 44.521 & 1.188 & 1.54700 & 44.799 & 1.260 \\
0.71500 & 41.551 & 2.122 & 1.11810 & 43.078 & 1.205 & 1.54700 & 47.147 & 1.280 \\
0.71600 & 41.651 & 1.348 & 1.13450 & 43.786 & 1.221 & 1.54900 & 43.025 & 1.202 \\
0.71600 & 40.606 & 1.337 & 1.14500 & 45.839 & 1.264 & 1.55100 & 45.404 & 1.514 \\
0.72500 & 45.584 & 1.279 & 1.14800 & 44.961 & 1.236 & 1.55450 & 47.598 & 1.415 \\
\bottomrule
\end{tabular}
\end{table*}
\begin{table*}[p]
\centering
\caption{Complete supplied GRB Hubble diagram (rows 133--264). Each group of three columns lists redshift, distance modulus, and its uncertainty. Rows follow the supplied file; no numerical values have been changed.}
\label{calibratedHd_full_cont}
\small
\begin{tabular}{rrr@{\hspace{14mm}}rrr@{\hspace{14mm}}rrr}
\toprule
$z$ & $\mu$ & $\sigma_\mu$ & $z$ & $\mu$ & $\sigma_\mu$ & $z$ & $\mu$ & $\sigma_\mu$ \\
\midrule
1.56300 & 42.867 & 2.179 & 2.10620 & 44.878 & 1.254 & 3.03630 & 46.793 & 1.309 \\
1.56700 & 44.707 & 1.241 & 2.14000 & 45.380 & 1.214 & 3.03800 & 45.563 & 1.470 \\
1.60000 & 45.536 & 1.469 & 2.14500 & 48.583 & 1.384 & 3.07500 & 47.664 & 1.321 \\
1.60435 & 47.765 & 1.377 & 2.19740 & 45.790 & 1.330 & 3.08400 & 46.092 & 1.368 \\
1.60800 & 48.200 & 1.431 & 2.19900 & 47.287 & 1.359 & 3.13900 & 45.995 & 1.486 \\
1.61300 & 45.735 & 1.211 & 2.20000 & 45.393 & 1.189 & 3.14000 & 47.004 & 1.282 \\
1.61900 & 45.102 & 1.240 & 2.20450 & 46.957 & 1.262 & 3.19800 & 46.678 & 1.389 \\
1.64030 & 45.910 & 1.223 & 2.22000 & 46.313 & 1.517 & 3.22000 & 45.977 & 1.309 \\
1.68600 & 45.156 & 1.212 & 2.24800 & 47.514 & 1.267 & 3.26000 & 47.358 & 1.258 \\
1.69190 & 43.637 & 1.369 & 2.26000 & 44.191 & 1.187 & 3.35500 & 46.022 & 1.301 \\
1.70000 & 46.245 & 1.212 & 2.27000 & 46.086 & 1.371 & 3.36000 & 48.926 & 1.351 \\
1.72700 & 45.360 & 1.350 & 2.27500 & 46.527 & 1.252 & 3.37000 & 48.144 & 1.531 \\
1.72800 & 44.509 & 1.193 & 2.30000 & 46.922 & 1.530 & 3.42000 & 48.082 & 1.283 \\
1.73600 & 45.682 & 1.279 & 2.30000 & 46.563 & 1.246 & 3.42400 & 47.800 & 1.331 \\
1.75800 & 45.756 & 1.264 & 2.32000 & 46.516 & 1.212 & 3.50000 & 46.135 & 1.193 \\
1.77000 & 43.965 & 1.194 & 2.33000 & 47.322 & 1.306 & 3.50300 & 49.108 & 1.267 \\
1.79800 & 43.924 & 1.276 & 2.33000 & 44.865 & 1.210 & 3.51000 & 46.824 & 1.300 \\
1.80000 & 46.407 & 1.241 & 2.34600 & 47.820 & 1.449 & 3.53280 & 47.488 & 1.201 \\
1.80000 & 45.119 & 1.231 & 2.40000 & 47.109 & 1.286 & 3.60000 & 47.163 & 1.387 \\
1.82200 & 46.209 & 1.394 & 2.40900 & 43.833 & 1.407 & 3.79600 & 48.146 & 1.229 \\
1.85800 & 46.022 & 1.533 & 2.43300 & 47.056 & 1.324 & 3.91000 & 47.623 & 1.396 \\
1.87400 & 46.633 & 1.266 & 2.46900 & 47.715 & 1.260 & 3.93000 & 48.344 & 1.539 \\
1.88360 & 47.239 & 1.610 & 2.48600 & 45.679 & 1.192 & 3.95600 & 47.504 & 1.330 \\
1.90000 & 46.537 & 1.398 & 2.48800 & 46.436 & 1.320 & 4.04200 & 47.039 & 1.237 \\
1.92000 & 49.255 & 1.367 & 2.51200 & 47.331 & 1.244 & 4.05590 & 49.190 & 1.632 \\
1.92290 & 45.737 & 1.419 & 2.53000 & 45.464 & 1.296 & 4.10900 & 47.109 & 1.248 \\
1.93700 & 43.519 & 1.222 & 2.58000 & 45.256 & 1.195 & 4.17450 & 48.495 & 1.282 \\
1.94920 & 47.565 & 1.453 & 2.59100 & 47.371 & 1.353 & 4.35000 & 48.415 & 1.436 \\
1.96800 & 47.522 & 1.660 & 2.61500 & 47.179 & 1.370 & 4.50000 & 47.313 & 1.571 \\
1.96850 & 45.313 & 1.249 & 2.65000 & 46.109 & 1.290 & 4.61800 & 47.372 & 1.321 \\
1.98000 & 45.295 & 1.273 & 2.67100 & 44.960 & 1.232 & 4.64970 & 49.829 & 1.435 \\
2.00600 & 46.788 & 1.698 & 2.68920 & 46.353 & 1.266 & 4.70700 & 47.637 & 1.230 \\
2.01000 & 45.112 & 1.186 & 2.70800 & 45.938 & 1.224 & 4.96100 & 48.297 & 1.382 \\
2.01300 & 48.084 & 1.269 & 2.73000 & 46.232 & 1.196 & 5.00000 & 48.148 & 1.281 \\
2.03690 & 44.805 & 1.187 & 2.75200 & 46.169 & 1.335 & 5.46000 & 47.902 & 1.335 \\
2.04000 & 44.907 & 1.216 & 2.77000 & 46.901 & 1.374 & 5.60000 & 48.341 & 1.258 \\
2.05000 & 48.147 & 1.756 & 2.77000 & 47.154 & 1.221 & 5.70000 & 49.846 & 1.324 \\
2.06000 & 46.078 & 1.285 & 2.82100 & 46.628 & 1.276 & 6.29500 & 51.140 & 1.645 \\
2.06270 & 46.599 & 1.300 & 2.82600 & 43.216 & 2.119 & 6.31800 & 48.662 & 1.287 \\
2.08580 & 45.449 & 1.337 & 2.83000 & 49.217 & 1.565 & 6.32000 & 49.401 & 1.334 \\
2.08700 & 45.714 & 1.228 & 2.89300 & 48.467 & 1.351 & 6.69500 & 50.840 & 1.665 \\
2.08800 & 46.708 & 1.244 & 2.89920 & 45.675 & 1.194 & 7.80000 & 50.522 & 1.340 \\
2.10000 & 44.418 & 1.200 & 2.93800 & 47.926 & 1.322 & 8.10000 & 50.130 & 1.313 \\
2.10600 & 47.096 & 1.211 & 3.00000 & 47.272 & 1.223 & 9.40000 & 50.384 & 1.316 \\
\bottomrule
\end{tabular}
\end{table*}
\clearpage
\end{document}